\documentclass[fleqn,usenatbib]{mnras}

\usepackage{newtxtext,newtxmath}

\usepackage[T1]{fontenc}

\DeclareRobustCommand{\VAN}[3]{#2}
\let\VANthebibliography\thebibliography
\def\thebibliography{\DeclareRobustCommand{\VAN}[3]{##3}\VANthebibliography}

\usepackage{graphicx}	
\usepackage{amsmath}	
\usepackage{subcaption}
\usepackage{cases}
\usepackage{multirow}
\usepackage{soul}
\usepackage[normalem]{ulem}

\newcommand{\Msun}{\,{\rm M}_{\odot}}

\newcommand{\Lsun}{\,{\rm L}_{\odot}}
\newcommand{\iso}[2]{\hbox{${}^{#1}{\rm #2}$}}
\newcommand{\stkout}[1]{\ifmmode\text{\sout{\ensuremath{#1}}}\else\sout{#1}\fi}

\title[$^{26}$Al production in AGB binary systems]{Aluminium-26 production in low- and intermediate-mass binary systems}

\author[Osborn et al.]{
Zara Osborn,$^{1,2}$\thanks{E-mail: zara.osborn@monash.edu}
Amanda I. Karakas,$^{1,2}$
Alex J. Kemp $^{3}$
and Robert G. Izzard$^{4}$
\\
$^{1}$School of Physics \& Astronomy, Monash University, Clayton 3800, VIC, Australia\\
$^{2}$Centre of Excellence for Astrophysics in Three Dimensions (ASTRO-3D), Melbourne, VIC, Australia\\
$^{3}$Institute of Astronomy, KU Leuven, Celestijnenlaan 200D, 3001 Leuven, Belgium \\
$^{4}$Astrophysics Research Group, University of Surrey, Guildford, Surrey GU2 7XH, UK\\
}

\date{Accepted XXX. Received YYY; in original form ZZZ}

\pubyear{2023}

\begin{document}
\label{firstpage}
\pagerange{\pageref{firstpage}--\pageref{lastpage}}
\maketitle

\begin{abstract}
Aluminium-26 is a radioactive isotope which can be synthesized within asymptotic giant branch (AGB) stars, primarily through hot bottom burning. Studies exploring \iso{26}Al production within AGB stars typically focus on single-stars; however, observations show that low- and intermediate-mass stars commonly exist in binaries. We use the binary population synthesis code \textsc{binary\_c} to explore the impact of binary evolution on \iso{26}Al yields at solar metallicity both within individual AGB stars and a low/intermediate-mass stellar population. We find the key stellar structural condition achieving most \iso{26}Al overproduction is for stars to enter the thermally-pulsing AGB (TP-AGB) phase with small cores relative to their total masses, allowing those stars to spend abnormally long times on the TP-AGB compared to single-stars of identical mass. Our population with a binary fraction of 0.75 has an \iso{26}Al weighted population yield increase of $25\%$ compared to our population of only single-stars. Stellar-models calculated from the Mt Stromlo/Monash Stellar Structure Program, which we use to test our results from \textsc{binary\_c} and closely examine the interior structure of the overproducing stars, support our \textsc{binary\_c} results only when the stellar envelope gains mass after core-He depletion. Stars which gain mass before core-He depletion still overproduce \iso{26}Al, but to a lesser extent. This introduces some physical uncertainty into our conclusions as 55\% of our \iso{26}Al overproducing stars gain envelope mass through stellar wind accretion onto pre-AGB objects. Our work highlights the need to consider binary influence on the production of \iso{26}Al.

\end{abstract}




\section{Introduction}

Aluminium-26 is a radioactive isotope ($\tau_{1/2} = 7.17 \times 10^5$ yrs) whose presence has caught the attention of researchers for being both a signature of ongoing Galactic nucleosynthesis \citep{Prantzos1996, Naya1998} and as an important heat source in the early solar system \citep{Lichtenberg2016}. It is estimated the early solar system had a \iso{26}Al/\iso{27}Al number ratio of $\approx 5.3 \times 10^{-5}$ \citep{Lee1976, Jacobsen2008, Mishra2014, Liu2019}, although its origin is uncertain \citep{Gaidos2009, Gaches2020, Parker2023}. Measurements of Galactic 1.809MeV $\gamma$-photons, which are attributed to the $\beta$-decay of \iso{26}Al, imply that the Galaxy currently contains $2.6-4.6\Msun$ of \iso{26}Al \citep{Naya1998}. Our paper investigates low- and intermediate-mass binary systems as a potential source of \iso{26}Al in the Galaxy.

While the majority of Galactic \iso{26}Al is typically attributed to massive stars \citep{Walter1989, Meynet1997, Knodlseder1999, Martin2009}, there are regions of 1.809MeV $\gamma$-photon emissions which suggest some Galactic \iso{26}Al originates from asymptotic giant branch (AGB) stars \citep{Naya1998}. \iso{26}Al production within AGB stars is also evidenced through measurements of \iso{26}Al/\iso{27}Al of up to $\sim0.1$ \citep{Nguyen2004, Lugaro2017} in presolar grains which are formed in stellar winds and indicates the parent stars' surface abundances at the time the material is ejected.

AGB stars are low- to intermediate-mass (${\sim} 0.8-9\Msun$) stars in their final nuclear burning phase. Within the AGB phase there are two sub-phases: the early-AGB (EAGB), which occurs between the end of core He-burning (CHe-B) and the onset of the first thermal pulse; and the thermally-pulsing AGB (TP-AGB), which is characterised by pulsing cycles driven by unstable shell He burning (see \citealp{Herwig2005} and \citealp{Karakas2014} for detailed reviews of the AGB phase). The production of \iso{26}Al within AGB stars relies mostly on proton capture onto \iso{25}Mg in H-burning regions at temperatures of at least $30$ million Kelvin \citep{Arnould1999}. 

Stellar models show that some \iso{26}Al is synthezised within the H-burning shell during the TP-AGB phase and transported to the stellar surface via third dredge up \citep{Forestini1991, vanRaai2008}. The third dredge ups describe the convective mixing of shell H- and He-burning products into the stellar envelope and potentially occurs after every thermal pulse cycle. However, \iso{26}Al has a relatively large neutron-capture cross-section \citep[$\approx 200$ mbarn,][]{deSmet2007}. This results in the destruction of most \iso{26}Al produced by the H-shell during convective shell He burning, before next third dredge up, due to the release of neutrons into the intershell. The efficiency of \iso{26}Al destruction via neutron-capture is highly correlated to the thermal pulse temperature, with close to $100\%$ of intershell \iso{26}Al destroyed at thermal pulse temperatures $>300 \times 10^6 \, \mathrm{K}$. These thermal pulse temperatures are achieved by solar metallicity stars of mass $\gtrsim 4 \Msun$ after the first few thermal pulses \citep{Mowlavi2000, Lugaro2008}. 

AGB stars of mass $\gtrsim 5\Msun$ synthesize \iso{26}Al more efficiently because their convective envelopes are massive enough to sustain H-burning temperatures ($\gtrsim 40 \times 10^6 \, \mathrm{K}$) at the base in a process known as hot-bottom burning \citep[HBB, ][]{Lattanzio1992, Boothroyd1992, Boothroyd1995}. As HBB does not provide a source of neutrons, \iso{26}Al survives within the stellar envelope. Therefore, the majority of \iso{26}Al ejected by AGB stellar models is synthesized via HBB \citep{Doherty2014, Karakas2016}. There have been multiple studies investigating \iso{26}Al production within single AGB stars \citep[e.g., ][]{Karakas2003,Ventura2011} and massive binaries \citep[e.g., ][]{Walter1989, Brinkman2019, Brinkman2021, Brinkman2023}, but the effect of binary stellar evolution on \iso{26}Al production by AGB stars remains unexplored. We aim to fill in this gap of knowledge and to study the affect of binary evolution on \iso{26}Al production. 

Binary stellar evolution can drastically alter a star's evolutionary pathway through a number of channels, including: Roche-Lobe overflow \citep[RLOF, ][]{Eggleton1983}, wind-Roche Lobe overflow \citep[WRLOF, ][]{Abate2013}, common envelope evolution \citep[CE, ][]{Paczynski1971}, and mergers (see \citealp{Iben1991} and \citealp{DeMarco2017} for detailed reviews of binary evolution). The probability of a binary interaction is highly dependent on the star's separation and their stellar radii. Because low- and intermediate-mass stars expand considerably on both the first giant branch and AGB, it is likely that HBB and \iso{26}Al production are especially vulnerable to binary influence. It is estimated that about 40\% of solar-like stars (${\sim}0.8-1.2\Msun$) have at least one stellar companion. In intermediate-mass (${\sim} 5-9\Msun$) stars companions are observed in $60-76$\% of systems \citep{Raghavan2010, Duchene2013, Moe2017}. Therefore, it is important to understand the influence of binary interactions on stellar evolution and the consequences on a stellar population.

Detailed AGB models are solutions to the equations of stellar structure along with other crucial input physics, e.g., reaction rates, opacities, equation of state, etc., however they are notoriously time-consuming especially when considering models with HBB \citep[see, for example, ][ for a comparison of stellar codes during the AGB phase]{Cinquegrana2022}. To investigate the production of \iso{26}Al in a stellar population of low- to intermediate-mass stars we instead turn to binary population synthesis.

Binary population synthesis relies on synthetic stellar models which are constructed from fitting formulae that approximate the results of detailed stellar models \citep[e.g. see ][]{Hurley2000, Hurley2002}. Population synthesis lets us quickly simulate many binary stellar models, permitting the exploration of vast regions of parameter-space in terms of the initial primary mass, secondary mass, and orbital period. \textsc{Binary\_c} \citep{Izzard2004, Izzard2006, Izzard2009} is a binary population synthesis code which models low- and intermediate-mass stellar evolution and nucleosynthesis, including
solving the nuclear networks involved in HBB \citep{Izzard2007}. 
\\ \\
Our goal is to investigate the consequences of binary influence on the yields of \iso{26}Al using simulations of stellar populations of low- and intermediate-mass stars. We do this using both binary population synthesis and detailed stellar models. Section \ref{sec:Method} details our methodology for calculating single- and binary-star models, and our methods for calculating the stellar yields. Section \ref{sec:Results} presents our results, we discuss our results in Section \ref{sec:Discussion} and we conclude in Section \ref{sec:Conclusion}. 

\section{Method}
\label{sec:Method}

This section outlines how we calculate our \iso{26}Al yields from AGB stars using both synthetic and detailed modelling techniques. We follow the methodology from \citet{Kemp2021} for our treatment of population statistics and normalisation. We use the acronyms defined in Table \ref{tab:Types} to describe various stellar objects.

\begin{table}
	\centering
	\caption{Acronyms from \citet{Hurley2000} used to identify stellar objects.}
	\label{tab:Types}
	\begin{tabular}{cc} 
        Acronym & Stellar evolutionary phase \\
		\hline
        MS & Main sequence \\
        HG & Hertzsprung gap \\
        GB & Giant branch \\
        CHeB & Core He burning \\
        EAGB & Early AGB \\
        TP-AGB & Thermally pulsing AGB \\
        naked-He & Naked helium \\
        naked-He HG & Naked helium Hertzsprung gap \\
        CO-WD & carbon-oxygen white dwarf \\
        NS & Neutron star \\
        BH & Black hole \\
		\hline
	\end{tabular}
\end{table}

\subsection{Stellar Synthetic Models}
\label{sec:PopSynth} 

In this work we are using the binary population synthesis code: \textsc{binary\_c V2.2.2}, hereafter referred to as the standard version of \textsc{binary\_c}, to calculate all synthetic models (single and binary-stars), as it is currently the only synthetic code available which models both AGB evolution and nucleosynthesis. \textsc{binary\_c} relies on fits to detailed stellar models from \citet{Pols1998} \citep[computed by ][]{Hurley2000, Hurley2002} and \citet{Karakas2002} \citep[computed by ][]{Izzard2004} to rapidly evolve isolated binary systems. It incorporates many improvements to areas of stellar physics including RLOF \citep{Claeys2014}, WROLF \citep{Abate2013, Abate2015}, tides \citep{Siess2013}, stellar rotation \citep{deMink2013}, binary stellar nucleosynthesis \citep{Izzard2018}, stellar lifetimes \citep{Schneider2014}, CE evolution \citep{Dewi2000, Wang2016}, and circumbinary disks \citep{Izzard2022}. 

We focus on a low-intermediate-mass population with initial masses between $0.95-8.5\Msun$ at solar metallicity ($Z=0.02$). AGB evolution and nucleosynthesis in \textsc{binary\_c} is calibrated for stellar masses up to $6.5\Msun$ using models from \citet{Karakas2002}, hereafter referred to as the \emph{Monash02} models. Stars of stellar mass $M \gtrsim 8.0 \Msun$ (these are massive stars, and do not experience the AGB) in \textsc{binary\_c} are calibrated to detailed models presented in \citet{Pols1998}, hereafter referred to as \emph{Pols98} models. For AGB stars of mass $6.5-8.0\Msun$ we use fits to \emph{Pols98} for stellar structure and fits to \emph{Monash02} at $6.5\Msun$ for AGB nucleosynthesis and handling of thermal pulses (since the \emph{Pols98} models skip over thermal pulses). There is also a smooth transition from the stellar structure fit to the \emph{Monash02} models to the \emph{Pols98} models in this region. Therefore TP-AGB evolution and nucleosynthesis in AGB stellar-models M $>6.5\Msun$ are uncalibrated. 

In our study, we extend the calibration of the AGB phase based on the newer Monash models up to $8\Msun$, making use of detailed stellar evolution calculations from \citet{Karakas2014_2}, \citet{Doherty2015}, and \citet{Karakas2016}. The stars with stellar structure described in \citet{Karakas2014_2} and nucleosynthesis calculated in \citet{Karakas2016} are hereafter referred to as the \emph{Monash16} models. Our modifications to \textsc{binary\_c} reduce the width of uncalibrated AGB evolution from $1.5\Msun$ to about $0.3\Msun$. This results in more realistic CO core masses (discussed in Section \ref{sub:CoreMassEAGB}), third dredge up parameters (Section \ref{sub:3DU}), and surface luminosities (\ref{sub:Luminosity}).

We note the slight difference in metallicity between the \emph{Monash02} (Z=0.02) and \emph{Monash16} (Z=0.014) models. This results in insignificant structural differences between the models (see Table 2 in \citet{Karakas2002} and Table 1 in \citet{Karakas2014_2} for comparison) with the largest difference being that the \emph{Monash16} models experience more thermal pulses. We do not anticipate that this will cause any significant change in our results since \iso{26}Al is primarily produced through HBB.

\subsubsection{Core mass during the EAGB and at the first thermal pulse}
\label{sub:CoreMassEAGB}

\begin{figure}
	\includegraphics[width=\columnwidth]{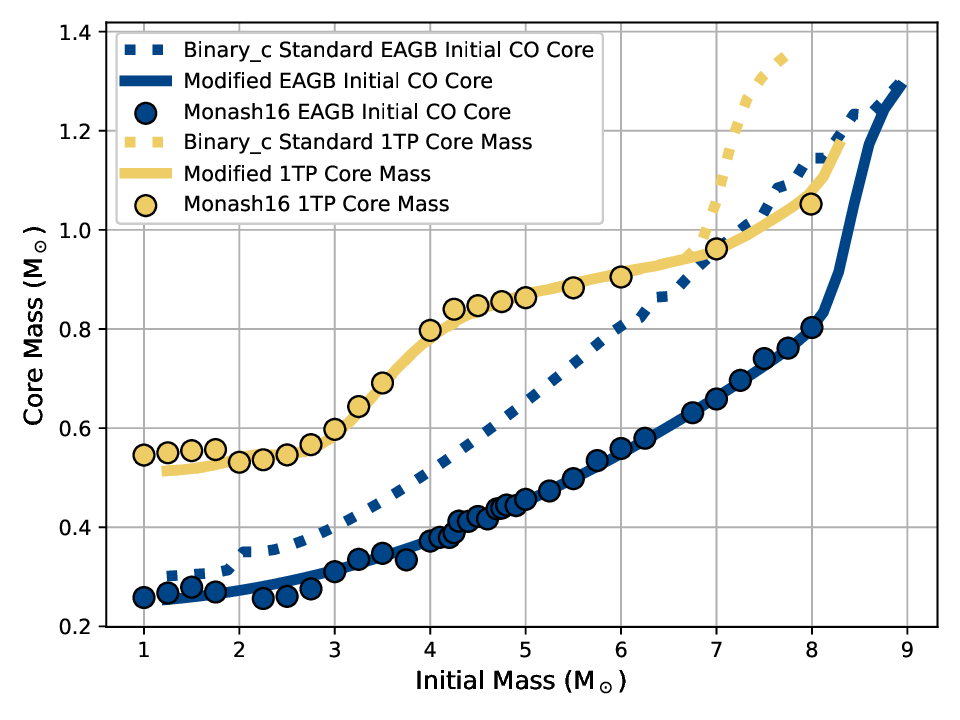}
    \caption{$M_{\rm {CO,BAGB}}$ (shown in yellow) and $M_{\rm {c,1TP}}$ (shown in blue) fits from the standard version of \textsc{binary\_c} (dotted lines), and our new fits (solid lines) to the \emph{Monash16} stellar-models (markers). The standard version of \textsc{binary\_c} uses fits to the \emph{Monash02} models to calculate $M_{\rm {CO,BAGB}}$ and $M_{\rm {c,1TP}}$ up to $6.5\Msun$ before smoothly transitioning to fits to the \emph{Pols98} models. We calibrate our new fits for $M_{\rm {CO,BAGB}}$ and $M_{\rm {c,1TP}}$ to the \emph{Monash16} models up to $8.0\Msun$, after which we smoothly transition to the \emph{Pols98} fits.}
    \label{fig:CoreMass}
\end{figure}

The core mass is one of the most important stellar parameters as it influences many stellar evolution variables such as radiated luminosity, radii, and the temperature at the base of the convective envelope. In the standard version of \textsc{binary\_c}, the CO core mass at the first thermal pulse ($M_{\rm{c,1TP}}$ in $\Msun$) is fit to \emph{Monash02} and \emph{Pols1998} models. A key difference between the \emph{Monash02} and \emph{Pols98} models is that, unlike the \emph{Pols98} models, the \emph{Monash02} models have no convective overshoot during the core H and core He-burning phases. The inclusion of overshoot results in the \emph{Pols98} cores being more massive than the \emph{Monash02} models for a given stellar mass. 

 The standard version of \textsc{binary\_c} smoothly transitions between the \emph{Monash02} and \emph{Pols98} fits near $6.5\Msun$, resulting in a steep increase in $M_{\rm{c,1TP}}$ as observed in Fig. \ref{fig:CoreMass}. The increase in core mass is not replicated by the \emph{Monash16} models. The more massive cores from the \emph{Pols98} fit result in increased temperatures at the base of the convective envelope which facilitate proton capture onto oxygen. This leads to oxygen depletion in AGB stars with an initial mass $>7.38\Msun$ and results in stars becoming carbon-rich as surface C/O $>1$, which is a behaviour not reflected by solar metallicity detailed stellar-models at this mass \citep{Siess2010, Doherty2014, Karakas2016}. Our solution is to extend the calibrated mass range of the $M_{\rm {c,1TP}}$ fitting formula presented in Eq (23) of \citet{Izzard2004} in the standard version of \textsc{binary\_c} to include the (non-overshooting) models of \citet{Karakas2014_2} and \citet{Doherty2015}:


\begin{equation}
    \label{eq:Mc1TP}
    M_{\rm{c,1TP}} = 
\begin{cases}
    \text{Eq. 23; \citet{Izzard2004}} & M_{\rm PostMS} \leq 6.88\Msun \\
    b_{\rm 10} + b_{\rm 11} M_{\rm PostMS} + b_{\rm 12} M_{\rm PostMS}^2 & 6.88 < M_{\rm PostMS} < 8.30 \Msun \\
    \text{Eq. 69; \citet{Hurley2000}} & M_{\rm PostMS} \gtrsim 8.30\Msun
\end{cases}
\end{equation}
\\
\noindent where $M_{\rm{c,1TP}}$ is in $\Msun$, $b_{\rm 10} = 1.227333$, $b_{\rm 11} = -0.176598$, $b_{\rm 12} = 0.019773$, and $M_{\rm PostMS}$ is the total mass of the star as it begins to cross the HG in $\Msun$. Eq. (69) from \citet{Hurley2000} is the fit for $M_{\rm{c,1TP}}$ to the \emph{Pols98} models. Eq. \ref{eq:Mc1TP} is valid for AGB stars of initial mass up to $9.8\Msun$ at $Z=0.02$, however we begin to smoothly transition to the \emph{Pols98} fit at $8.30\Msun$ where the synthetic stars non-degenerately ignite the CO core in both the standard and our modified versions of \textsc{binary\_c}. 

Fig. \ref{fig:CoreMass} shows the fits for $M_{\rm{c,1TP}}$ in both the standard and our modified versions of \textsc{binary\_c} with the $M_{\rm{c,1TP}}$ calculated by our \emph{Monash16} models. Fig. \ref{fig:CoreMass} shows our modified fit is in better agreement with our \emph{Monash16} models than the fit in the standard version of \textsc{binary\_c} within the initial mass range of $6.5-8\Msun$.

Fig. \ref{fig:CoreMass} shows that when $M \gtrsim 7\Msun$ our re-fitted $M_{\rm {c,1TP}}$ is less massive than the CO core mass at the beginning of the EAGB (hereafter denoted as $M_{\rm{CO,BAGB}}$). A condition for \textsc{binary\_c} to successfully evolve a star through the AGB phase is for $M_{\rm{c,1TP}} > M_{\rm{CO,BAGB}}$. In the standard version of \textsc{binary\_c} the CO core mass is initially calculated at the beginning of the EAGB where $M_{\rm{CO,BAGB}}$ is fit to the \emph{Pols98} models. The CO core mass for stars with $M<6.5\Msun$ then transition to using the \emph{Monash02} fit at the first thermal pulse. To rectify this issue $M_{\rm{CO,BAGB}}$ was refitted to the \emph{Monash16} models via:

\begin{equation}
    \label{eq:EAGB_COinit}
    M_{\rm{CO,BAGB}} = b_{\rm 20} + b_{\rm 21}M_{\rm PostMS} + b_{\rm 22}M_{\rm PostMS}^2,
\end{equation}
\\ 
\noindent where $b_{\rm 20} = 0.247711$, $b_{\rm 21} = -0.006649$, and $b_{\rm 22} = 0.009530$. Eq. \ref{eq:EAGB_COinit} is valid for masses $1-8\Msun$ at $Z=0.02$. Our resulting fit for $M_{\rm{CO,BAGB}}$, the fit from the standard version of \textsc{binary\_c}, and $M_{\rm{CO,BAGB}}$ calculated by our \emph{Monash16} models are presented in Fig. \ref{fig:CoreMass}. In our modified version of \textsc{binary\_c} we use the existing CO core growth algorithm present in the standard version of \textsc{binary\_c} during the EAGB described by Eq. (33) in \citet{Hurley2000}. To accommodate the transition to the massive star regime at masses exceeding $8.30\Msun$, our modified version of \textsc{binary\_c} smoothly transitions our $M_{\rm{CO,BAGB}}$ fit to the \emph{Pols98} fit mimicking the treatment the standard version of \textsc{binary\_c} uses when fitting $M_{\rm{c,1TP}}$.

\subsubsection{Maximum third dredge up parameter}
\label{sub:3DU}
The third dredge-up parameter, $\lambda$, describes the efficiency of the third dredge up \citep{Karakas2002}:

\begin{equation}
    \lambda = \frac{\Delta M_{\rm{dredge}}}{\Delta M_{\rm c}},
    \label{eq:LamMax}
\end{equation}

\noindent where $\Delta M_{\rm{dredge}}$ is the mass of material which is dredged up and $\Delta M_{\rm c}$ is the core mass increase during the interpulse period. The standard version of \textsc{binary\_c} calibrates the third dredge up maximum efficiency, $\lambda_{\rm{max}}$, up to a maximum initial stellar mass of $6.5\Msun$. We refit the parameters $b_{\rm i}$ for $\lambda_{\rm{max}}$ from Eq. (6) in \citet{Karakas2002} using the \emph{Monash16} models up to $8\Msun$. Our modified fit is:

\begin{equation}
    \label{eq:lambdaMax}
    \lambda_{\rm{max}} = \frac{b_{\rm 31} + b_{\rm 32}M_{\rm PostMS} + b_{\rm 33}M_{\rm PostMS}^3}{1 + b_{\rm 34}M_{\rm PostMS}^3}
\end{equation}
\\
\noindent where $b_{\rm 31} = -0.371188$, $b_{\rm 32} = 0.417241$, $b_{\rm 33} = 0.021788$, and $b_{\rm 34} = 0.028935$. Eq. \ref{eq:lambdaMax} is valid for mass $1-8\Msun$ at $Z=0.02$.

\subsubsection{Luminosity during TP-AGB}
\label{sub:Luminosity}
Table A.1 of \citet{Izzard2006} is the updated table used by Eq. (29) of \citet{Izzard2004} to calculate the surface luminosity during the TP-AGB phase and is calibrated up to $6\Msun$. Using the \emph{Monash16} models, we have expanded the table to include stars up to $8\Msun$, as shown in Table \ref{tab:Luminosity}.

\begin{table}
	\centering
	\caption{Table A.1, of \citet{Izzard2006} updated to include up to $8\Msun$ stars at $Z = 0.02$. $N_{\rm TO}$ describes the rise in luminosity during the first few thermal pulses, and $f_{\rm{turnon,min}}$ is the minimum modulation factor of the peak luminosity \citep[see Eq. A.6 of ][]{Izzard2006}. }
	\label{tab:Luminosity}
	\begin{tabular}{lcr} 
		\hline
		$M$ ($\Msun$) & $N_{\rm{TO}}$ & $f_{\rm{turnon,min}}$\\
		\hline
		1 & 11 & 0.4 \\
		2 & 12 & 0.35 \\
		3 & 15 & 0.35 \\
        4 & 14 & 0.4 \\
        5 & 24 & 0.4 \\
        6 & 17 & 0.4 \\
        7 & 19 & 0.5 \\
        8 & 30 & 0.8 \\
		\hline
	\end{tabular}
\end{table}

\subsubsection{Temperature at the base of the convective envelope during the TP-AGB}
\label{sub:hbbtmax}
\begin{figure}
	\includegraphics[width=\columnwidth]{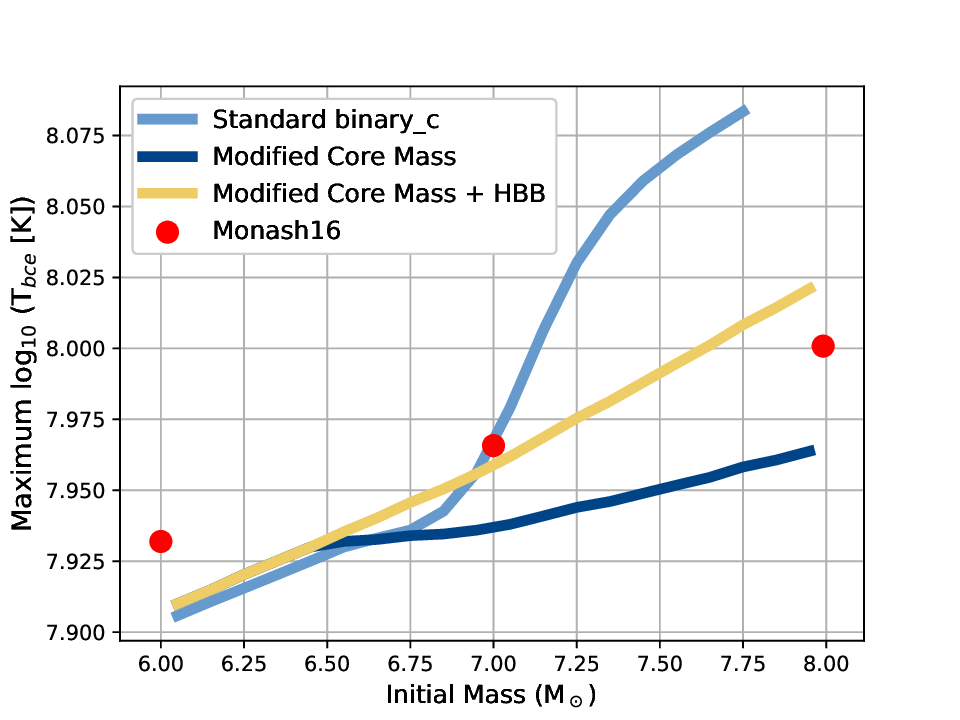}
    \caption{Maximum $T_{\rm bce}$ achieved by single-stars modelled using the standard version of \textsc{binary\_c} (dark blue), our modified version of \textsc{binary\_c} using the CO core masses described in Section \ref{sub:CoreMassEAGB} (light blue), and our modified version of \textsc{binary\_c} including our modifications to both the CO core mass and maximum $T_{\rm bce}$. We also show the maximum $T_{\rm bce}$ from stars calculated by the \emph{Monash16} models (red markers). Our modifications to the CO core mass combined with our extrapolation of the maximum $T_{\rm bce}$ fit results in the maximum $T_{\rm bce}$ from our stars calculated from our modified version of \textsc{binary\_c} being in better agreement to the \emph{Monash16} stellar-models, compared to our stars modelled using the standard version of \textsc{binary\_c}.}
    \label{fig:hbbtmax}
\end{figure}

The temperature at the base of the convective envelope $T_{\rm bce}$ governs the rate of HBB and hence \iso{26}Al production. The standard version of \textsc{binary\_c} uses a fit to the maximum temperature of a star over its lifetime and regulates it over the TP-AGB. The maximum $T_{\rm bce}$ throughout the lifetime of the star is described by Eq. 37 and 38 in \citet{Izzard2004} as:

\begin{equation}
    \label{eq:hbbtmax_Iz2004_1}
    \rm{log_{10}}(T_{\rm bce, max}) = {\rm min}(6.0379 + a_{\rm 37} M_{\rm env,1TP} + B(\zeta, M_{\rm c,1TP}), 7.95)
\end{equation}

\noindent where, $\zeta = 0$ at metallicity 0.02, $a_{\rm 37}$ is a constant, and

\begin{equation}
\label{eq:hbbtmax_Iz2004_2}
    B(\zeta, M_{\rm c,1TP}) = (a_{\rm 38} \zeta^2 + b_{\rm 38} \zeta + c_{\rm 38}) \times \left[ 1+d_{\rm 38} M_{\rm c,1TP} + e_{\rm 38} M_{\rm c,1TP}^2 \right]
\end{equation}

\noindent where $a_{\rm 38} ... e_{\rm 38}$ are constants. The maximum limit of $\rm{log_{10}}(T_{\rm bce, max}) = 7.95$ no longer exists in the standard version of \textsc{binary\_c}. See Sec. 3.3.4 of \citet{Izzard2004} for more detail about how \textsc{binary\_c} models temperatures at the base of the convective envelope.

The standard version of \textsc{binary\_c} calibrates the maximum $T_{\rm bce}$ fit up to $6.5\Msun$ and restricts parameters input into the fit to $M_{\rm env,1TP} = 5.5\Msun$ and $M_{\rm c,1TP} = 1.38\Msun$. In our modified version of \textsc{binary\_c}, our reduced core mass also leads to a reduction in the maximum $T_{\rm bce}$ leading to a notable reduction in \iso{26}Al production. To compensate we allow Eq. \ref{eq:hbbtmax_Iz2004_1} and \ref{eq:hbbtmax_Iz2004_2} to extrapolate to $8\Msun$ by changing the maximum $M_{\rm env,1TP}$ and $M_{\rm c,1TP}$ to $6.83\Msun$ and $M_{\rm c,1TP} = 1.01\Msun$ respectively. The new $M_{\rm c,1TP}$ and $M_{\rm c,1TP}$ limits are based on the $M_{\rm c,1TP}$ and $M_{\rm c,1TP}$ of a single $8\Msun$ synthetic model using the CO core mass algorithm described in Section \ref{sub:CoreMassEAGB}. 

Fig. \ref{fig:hbbtmax} shows the result of our modifications to the CO core mass and our extrapolation to Eq. (37) in \citet{Izzard2004} on the maximum $T_{\rm bce}$ achieved by our modelled stars over their lifetime, compared with the standard version of \textsc{binary\_c} and the \emph{Monash16} models. Fig. \ref{fig:hbbtmax} shows our reduction of the CO core during the TP-AGB also reduced the maximum $T_{\rm bce}$ below that of the stars from the \emph{Monash16} models. Fig. \ref{fig:hbbtmax} shows our extrapolation of the maximum $T_{\rm bce}$ fit in combination with our modified core mass, allows stars modelled using our modified version of \textsc{binary\_c} to experience a trend of increasing maximum $T_{\rm bce}$ with initial mass, similar to the \emph{Monash16} models. The maximum $T_{\rm bce}$ calculated using our modified version of \textsc{binary\_c} are also in better agreement to the \emph{Monash16} models than the standard version of \textsc{binary\_c}. 

\subsubsection{\textsc{binary\_c} input parameters}

%
\begin{table*}
	\centering
	\caption{Input physics and parameters of our \textsc{binary\_c} grids. Parameters not listed are set to the \textsc{binary\_c} V2.2.2 defaults described in \citet{Claeys2014}.}
	\label{tab:Inputs}
	\begin{tabular}{lr} 
		\hline
		\textbf{Parameter/Physics} & \textbf{Standard} \\
		\hline
		Primary-star initial mass range, $M_{\rm{1,0}}$ & $0.95 - 8.5\Msun$ \\
        $M_{\rm{1,0}}$ sampling distribution & Uniform \\
        Secondary-star initial mass range, $M_{\rm{2,0}}$ & $0.1\Msun$ - $M_{\rm{1,0}}$ \\
        $M_{\rm{2,0}}$ sampling distribution & Uniform \\
		Initial orbital period, $p_{\rm 0}$ & $1.0$ - $10^6$ days \\
        $p_{\rm 0}$ Sampling distribution & Log-uniform \\
        Metallicity, $Z$ & 0.02 \\
        Simulation time & $15$ Gyr \\
        Initial eccentricity & 0.0 \\
        Initial stellar rotation & 0.0 \\
        CE efficiency parameter, $\alpha$ & 1.0 \\
        CE binding energy parameter, $\lambda_{\rm BE}$ & \citet{Dewi2000} \\
        Wind angular momentum loss & Spherically symmetric \\
        RLOF angular momentum transfer model & Conservative \\
        Non-conservative angular momentum loss & Isotropic \\
        Chandrasekhar mass & $1.38\Msun$ \\
        TP-AGB upper mass limit & $8.35\Msun$ \\
        AGB core/radius/luminosity algorithms & \citet{Karakas2002} \\
        Initial chemical abundance & \citet{Anders1989} \\
        Core He-burning stellar wind & Off \\
        TP-AGB stellar wind & \citet{Vassiliadis1993} \\
        RLOF method & \citet{Claeys2014} \\
        WRLOF method & $q$-dependent \citep{Abate2013} \\
		\hline
	\end{tabular}
\end{table*}

A grid of binary stellar-models were generated using our modified version of \textsc{binary\_c}. Our input parameters and prescriptions are summarized in Table \ref{tab:Inputs} and are based off the input parameters from \citet{Kemp2021}. Other model parameters are set to the \textsc{binary\_c} V2.2.2 defaults, most of which are described in \citet{Claeys2014}. A complete list of model parameters may be obtained upon request from the author.

The various AGB algorithms, initial chemical abundances, and TP-AGB wind prescription presented in Table \ref{tab:Inputs} are chosen to be the same as the prescriptions used to initially calibrate the \textsc{binary\_c} stellar-models to \emph{Monash02} in \citet{Izzard2004}. We use the CE binding energy parameter, $\lambda_{\rm BE}$, described by \citet{Dewi2000} as it allows $\lambda_{\rm BE}$ to change with stellar evolutionary phase. The RLOF and WRLOF prescriptions presented in Table \ref{tab:Inputs} were chosen based on their common usage within other studies \citep[][]{Izzard2018, Kemp2021} and are considered improvements to the default prescriptions. 

There is a threshold within \textsc{binary\_c} which describes the mass below which the TP-AGB prescriptions for radius, luminosity, and nucleosynthsis are used (shown as "TP-AGB upper mass limit in Table \ref{tab:Inputs}). By default it was set to $8.0\Msun$. In the standard version of \textsc{binary\_c} non-degenerate carbon ignition occurs in stars of masses $>7.64\Msun$ (although modelled stars up to $8.00\Msun$ experience the TP-AGB). Our modification to $M_{\rm c,1TP}$ described in Eq. \ref{eq:Mc1TP} results in less massive cores at the first thermal pulse, as shown in Fig. \ref{fig:CoreMass}, and non-degenerate carbon ignition occurring at masses $>8.30\Msun$. We increase the upper mass limit for the TP-AGB prescriptions to $8.35\Msun$ to accommodate.

\subsection{Stellar yields}
\label{sec:StellarYields}

We define "stellar yield" to be the total mass ejected of a specific isotope by a star over its lifetime. AGB stars eject matter into the interstellar medium via stellar winds, making their stellar yields dependent on their surface abundances. We discuss how stellar yields are calculated from our synthetic models in both single and binary-star cases. 

\subsubsection{Single-star stellar yields}
The stellar yields of single-stars are calculated via Eq. \ref{eq:SinYield} \citep{Karakas2010}:

\begin{equation}
    \label{eq:SinYield}
    y_{\rm k} = \int_{0}^{\tau} X_{k}(t) \frac{{\rm d}M}{{\rm d} t} {\rm d}t,
\end{equation}
\\
\noindent where $y_{k}$ is the yield of species $k$ in $\Msun$, $X_{k}$ is the surface mass fraction of species $k$ at time $t$, and $\frac{{\rm d}M}{{\rm d}t}$ is the mass loss rate at time $t$.

\subsubsection{Binary-star stellar yields}
Due to the complex nature of binary evolution, binary stellar yields require a more elaborate calculation than stellar yields from single-stars. The calculation of binary stellar yields includes treatments for the following four scenarios unique to binary stellar physics: (i) both stars undergoing mass loss, (ii) mass transfer, (iii) CE/mergers, and (iv) novae/supernovae.
\\ \\
\emph{(i) Both stars undergoing mass loss.}
When both stars are losing mass it is assumed that there is no mass transfer. Eq. \ref{eq:SinYield} is used to calculate the stellar yield from each individual star.
\\ \\
\emph{(ii) Mass transfer.}
During a mass transfer event such as RLOF or stellar wind accretion, all ejected mass is assumed to originate from the donating star. The net mass loss from the system during each time step is calculated by taking the difference between the mass lost from the donor star, and the mass gained by the accreting star.
\\ \\
\emph{(iii) CE/merger.}
In the event of a CE, the envelope is either ejected or the stars merge. If the envelope is ejected the stellar yield contribution is calculated using Eq. \ref{eq:SinYield} using the surface abundances of the overflowing star initially prior to the CE. If a merger occurs, all of the mass ejected from the post-merger object are taken to originate from the primary. Eq. \ref{eq:SinYield} is used to calculate the yield contribution from a post-merger object. If some of the envelope is ejected during a merger, we determine the mass ejected from each individual star from the following scenarios:

\begin{itemize}
    \item{\emph{Merger with a remnant.} Remnants are not considered to contribute to stellar yield. All mass loss from the system is assumed to originate from the non-remnant star. Nucleosynthesis resulting from mergers between two remnants (e.g., a double-degenerate Type Ia supernovae) are not considered in our stellar yield calculation.}
    \item{\emph{Merger with one TP-AGB star.} The radius of the envelope of the TP-AGB star is significantly larger than the radius of its core. Therefore the material in the envelope is considered to be bound relatively loosely and more likely to be ejected during a merger. Here we assume the TP-AGB star ejects its entire envelope before the other star contributes. For example if a TP-AGB star with a $0.6\Msun$ envelope merges with a MS star with a $1\Msun$ envelope and $0.8\Msun$ is ejected from the system, it is assumed that $0.6\Msun$ originates from the TP-AGB star and $0.2\Msun$ from the MS star.}
    \item{\emph{Merger between two non/both TP-AGB stars.} It is assumed that both stars eject material according to the ratio of their envelope masses. For example, should a star with $0.6\Msun$ envelope mass merge with a star that has $1.0\Msun$ envelope mass, and $0.8\Msun$ is ejected from the system, then it is assumed that $\frac{0.6}{0.6+1.0}\times 0.8=0.3\Msun$ of the total $0.8\Msun$ originates from the primary-star.}
\end{itemize}

\emph{(iv) Novae/supernovae.}
Binary evolution may cause stars (that would normally not explode when single) to explode. We do not consider yield contributions from novae and supernovae in this work, only stellar winds.

\subsection{Weighted population stellar yields}
\label{sec:PopYields}
The initial conditions for stellar-models in \textsc{binary\_c} are sampled from a grid of various independent initial conditions and are then allowed to evolve. We use a 3D grid: $M_{\rm1,0} \times M_{\rm 2,0} \times p_{\rm 0}$, which are sampled as described in Table \ref{tab:Inputs}. Grids of stellar-models in \textsc{binary\_c} do not reflect the physical birth distributions of a stellar population. To understand of the influence of binary evolution on a low- and intermediate-mass stellar population, we need to correct for this discrepancy by weighting the individual stellar yields by the theoretical birth mass probability distributions, $\pi$, of the stellar population. The weighting algorithm is based on the algorithm presented in \citet{Broekgaarden2019}, see also \citet{Kemp2021}. The normalised weighting factors (in units of $\Msun$ per $\Msun$ of star-forming material available to our population, denoted as $\Msun/{\rm M_{\odot,SFM}}$) are:

\begin{equation}
    \label{eq:NWeightS}
        w_{{\rm s},i} = (1-f_{\rm{b}}) \frac{w_{\rm{m}}}{n_{{\rm s}}}\frac{\pi(\mathbf{x}_{{\rm s},i})}{\xi(\mathbf{x}_{{\rm s},i})}, \\
\end{equation}
\\

\noindent for our single-star models, and

\begin{equation}
    \label{eq:NWeightB}
        w_{{\rm b},i} = f_{\rm{b}} \frac{w_{\rm{m}}}{n_{\rm{b}}}\frac{\pi(\mathbf{x}_{{\rm b},i})}{\xi(\mathbf{x}_{{\rm b},i})},
\end{equation}
\\

\noindent for our binary-star models where $f_{\rm b}$ is the binary fraction of our population, $\pi(\mathbf{x}_{{\rm s},i})$ and $\pi(\mathbf{x}_{{\rm b},i})$ are the theoretical birth probability distributions of the single and binary portions of the population respectively for system $i$, and $\xi(\mathbf{x}_{{\rm s},i})$ and $\xi(\mathbf{x}_{{\rm b},i})$ are the \textsc{binary\_c} grid sample probability distributions, $w_{\rm{m}}$ is a normalising factor which describes the number of stellar systems per ${\rm M_{\odot,SFM}}$, and $n_{\rm s}$ and $n_{\rm b}$ are the numbers of single and binary systems respectively sampled in \textsc{binary\_c}.

The parameters describing the initial conditions of a binary system are: the initial primary-star mass $M_{\rm {1,0}}$, the initial secondary star mass $M_{\rm {2,0}}$, and the initial orbital period $p_{\rm 0}$, and they are assumed to be independent of one another. We write the birth mass distribution of the binary systems as the product of the probability distribution of each parameter,

\begin{equation}
    \pi(\mathbf{x}_{\rm{b},i}) = P_{\rm 1}(M_{\rm{1,0}}) P_{\rm 2}(M_{\rm{1,0}}, M_{\rm{2,0}}) P_{\rm p}(p_{\rm 0}),
\end{equation}
\\

\noindent where $P_{\rm 1}$ is the initial mass function described by \citet{Kroupa2002}, $P_{\rm 2}$ is the birth mass probability of the secondary star masses, taken to be uniform between $0.1M_{\rm{1,0}}$ and $M_{\rm{1,0}}$, and $P_{\rm p}$ is the probability distribution of the initial orbital period, taken to be log-uniform between $1$ and $10^6$ days (see Table \ref{tab:Inputs}). For the single-star portion, like the binary primary-stars, the birth mass probability distribution was taken from the initial mass function described in \citet{Kroupa2002}. The repository \url{https://github.com/keflavich/imf} was utilised for calculations involving the initial mass function.

Equations \ref{eq:NWeightS} and \ref{eq:NWeightB} can then weight any stellar output, such as the stellar yield, to give us the weighted result for per $M_{\odot,{\rm SFM}}$ for system $i$. The weighted stellar yields produced by the whole single (denoted as $y_{k,\rm{sPop}}$ in units of $\Msun/{\rm M_{\odot,SFM}}$) and binary (denoted as $y_{k,\rm{bPop}}$ in units of $\Msun/{\rm M_{\odot,SFM}}$) portions of the stellar population are:

\begin{equation}
    \label{eq:WYields}
    y_{k,{\rm sPop}} = \sum_{i={\rm 1}}^{n_{\rm s}} w_{{\rm s},i} \times y_{k,{\rm s},i}, 
\end{equation}
\\
\noindent and

\begin{equation}
    \label{eq:WYieldsB}
    y_{k,{\rm bPop}} = \sum_{i={\rm 1}}^{n_{\rm b}} w_{{\rm b},i} \times y_{k,{\rm b},i},
\end{equation}
\\

\noindent where $y_{k,{\rm s},i}$ in $\Msun$ is the stellar yield of nuclide $k$ produced by single-star model $i$ of the total ${n_{\rm s}}$ number of single-star models, and $y_{k,{\rm b},i}$ in $\Msun$ is as $y_{k,{\rm s}}$ for the binary systems. $y_{k,{\rm b},i}$ describes the stellar yield from either the primary, secondary, or both. Finally, the total weighted stellar yield of nuclide $k$ produced by the low- and intermediate-mass stellar population, $y_{k,{\rm pop}}$ in $\Msun/{\rm M_{\odot,SFM}}$, is

\begin{equation}
    \label{eq:PopYieldsTot}
    y_{k,{\rm pop}} = y_{k,{\rm sPop}} + y_{k,{\rm bPop}}.
\end{equation}


Our weighted \iso{26}Al stellar population yields are calculated using Equations \ref{eq:NWeightS}  to \ref{eq:WYields} for populations with binary fractions 0, 0.25, 0.50, 0.75 and 1.0, using a $80 \times 80 \times 80$ grid in our modified version of \textsc{binary\_c}.

\subsection{Detailed stellar evolution models}
\label{sec:MonashModels}

A limitation of synthetic binary stellar models is that they rely on approximations to detailed single-star models. To mitigate this limitation, we employ detailed evolutionary modelling to test the key evolutionary channels identified using our population-synthesis results. In Section \ref{sec:weightedYields} we find that the inclusion of binaries in our population model results in more \iso{26}Al being ejected than our population of only single-stars. In Section \ref{sec:IMS_Results} we find some synthetic models responsible for \iso{26}Al overproduction enter the TP-AGB with relatively small cores for their total mass. Our goal is to test if we can replicate these conditions which allow for \iso{26}Al overproduction with detailed models and examine the consequences on the stellar structure and yields.

The Mt Stromlo/Monash Stellar Structure Program (hereafter referred to as the Monash code), as described in \citet{Karakas2016} and references therein, is used to closely examine the stellar structure of stars being influenced by binary evolution. The synthetically modelled stars overproduce \iso{26}Al when they enter the TP-AGB with a relatively massive envelope and small core for their total mass (more detail is presented in Section \ref{sub:Synth}). We replicate binary conditions within the Monash code by mimicking a merger with a MS star by increasing the envelope mass of the evolving star whilst holding the core mass constant. 

We evolve two stars using the Monash code both with an initial mass of $5\Msun$ and solar metallicity, $Z=0.014$ \citep{Asplund2009}, from the pre-MS. The input physics for the single $5\Msun$ star are as described in \citet{Karakas2016}, including mass-loss prescriptions. Table \ref{tab:MonashDump} shows the conditions of our single $5\Msun$ star when we begin adding mass to the envelope. We evolve one $5\Msun$ star until it crosses the HG but we halt its evolution before it ascends the GB. We increase the envelope mass by $1\Msun$, we turn on mass loss which allows the star to relax into hydrostatic and thermal equilibrium, and then increase the envelope mass by an additional $1\Msun$ and allow the star to relax once again. We allow the star to evolve until stellar winds during the TP-AGB reduce the total mass of the star to $3.78\Msun$ where subsequent models fail to converge. During the GB we use the mass loss prescription described in \citet{Reimers1975} with $\eta = 0.455$ \citep{McDonald2015}. We increase the envelope mass in $1\Msun$ increments as it allows us to rapidly add $2-3\Msun$ while also allowing the models to adjust and converge. A total mass of $7\Msun$ was the maximum mass achieved using this method as models of more massive stars failed to converge. We hereafter refer to this model as our $5+2\Msun$ star. 

\begin{table*}
	\centering
	\caption{Stellar conditions of our single $5\Msun$ star modelled using the Monash code just before we add extra mass to the stellar envelope. The time is the age of the star, $M_{\rm c}$ is the H-exhausted core mass, $M_{\rm env}$ is the envelope mass, $T_{\rm eff}$ is the effective temperature, and $L$ is the surface luminosity.}
	\label{tab:MonashDump}
	\begin{tabular}{l|ccccc} 
		\hline
		Model & Time (yr) & $M_{\rm c}$ ($\Msun$) & $M_{\rm env}$ ($\Msun$) & ${\rm log_{\rm 10}} (T_{\rm eff} \, [\mathrm{K}])$ & ${\rm log_{\rm 10}} (L \, [\Lsun]))$ \\ 
		\hline
        $5+2\Msun$ & $8.224 \times 10^7$ & 0.773 & 4.227 & 3.708 & 2.608 \\
        $5+3\Msun$ & $1.061 \times 10^8$ & 1.013 & 3.987 & 3.641 & 3.154 \\
        \hline
	\end{tabular}
\end{table*}

We evolve the other $5\Msun$ star until the beginning of the EAGB. During the AGB we use the mass loss prescription described in \citet{Vassiliadis1993}. We then halt the evolution of the star, increase the envelope mass by $1\Msun$, and allow the star to relax into hydrostatic and thermal equilibrium. We repeat this process two additional times resulting in a star with a total mass of $8\Msun$. We hereafter refer to this model as our $5+3\Msun$ star. We then allow our $5+3\Msun$ star to evolve through 105 thermal pulses. We use a $5+3\Msun$ binary model as many of our \iso{26}Al overproducing intermediate-mass synthetic models enter the TP-AGB with a mass $\approx 8\Msun$ (see Section \ref{sec:IMS_Results}).

\section{Results}
\label{sec:Results}

This section presents our results of our synthetic models described in Section \ref{sub:Synth}. We then compare our results from our synthetic models to our detailed models described in Section \ref{sub:Monash}.

\subsection{Synthetic model results}
\label{sub:Synth}
Here we present the results from our synthetic models calculated using our modified version of \textsc{binary\_c}. We first examine the \iso{26}Al yields from our single-star models, and then examine our stellar populations. From our population models we investigate the stellar conditions which influence the production of \iso{26}Al. 

\subsubsection{Single-star yields}

\begin{figure}
	\includegraphics[width=\columnwidth]{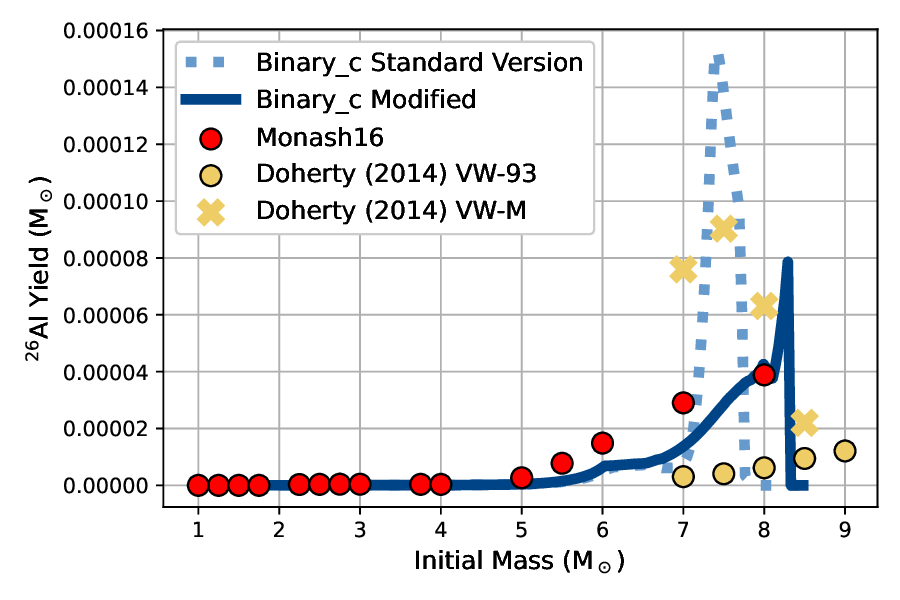}
    \caption{\iso{26}Al yields as calculated from the standard (light blue) and modified (dark blue) versions of \textsc{binary\_c}, the detailed \emph{Monash16} ($Z=0.014$, red markers) models, and \citet{Doherty2014} ($Z=0.02$, yellow markers) models. All models use the stellar wind prescription described in \citet{Vassiliadis1993}, with the superwind beginning once the radial pulsation period reaches 500 days, with the exception of the \citet{Doherty2014} models notated as VW-M (yellow crosses) where the superwind begins once the radial pulsation period reaches 850 days. AGB evolution in \textsc{binary\_c} is calibrated up to $6.5\Msun$ in the standard and $8.0$ in our modified versions. Uncalibrated AGB stars experience more massive CO-cores (see Fig. \ref{fig:CoreMass}) and increased HBB temperatures (see Fig. \ref{fig:hbbtmax}) leading to large spikes of \iso{26}Al production. Our modifications to \textsc{binary\_c} reduce the \iso{26}Al peak yield from $1.15 \times 10^{-4}\Msun$ to $7.86 \times 10^{-5}\Msun$. Overall, our modifications to \textsc{binary\_c} result in \iso{26}Al yields in better agreement to the stars from the \emph{Monash16} models.}
    \label{fig:Al26Single}
\end{figure}


We first calculate the \iso{26}Al stellar yields from our single-stars modelled by our modified version of \textsc{binary\_c}. Our goal is to verify that our changes described in Section \ref{sec:PopSynth} improve on the \iso{26}Al yields from stars modelled from the standard version of \textsc{binary\_c} and compared to stars from the \emph{Monash16} and \citet{Doherty2014} detailed models. Fig. \ref{fig:Al26Single} shows our \iso{26}Al yields from single-stars modelled by the standard and modified versions of \textsc{binary\_c} with initial masses $0.95-8.5\Msun$ along with stellar-yields from the \emph{Monash16} and \citet{Doherty2014} ($Z=0.02$, and with the superwind beginning at radial pulsation periods of 500 days and 850 days) models.

Fig. \ref{fig:Al26Single} shows a sharp spike in the \iso{26}Al yields for stars of initial mass $M > 6.5\Msun$ from the standard version of \textsc{binary\_c}, which is not reflected by the stellar-yields from \citet{Doherty2014} or \emph{Monash16}. The spike is caused by the transition from the \emph{Monash02} to the \emph{Pols98} models near initial mass $6.5\Msun$ and the mismatch between the stellar structure calibrated using the \emph{Pols98} models, and the thermal pulses AGB nucleosynthesis calibrated using \emph{Monash02} models. The \iso{26}Al yield peaks at $7.42\Msun$ with a \iso{26}Al yield of $1.15 \times 10^{-4} \Msun$. For comparison the $7.0\Msun$ star from the Monash code has an \iso{26}Al yield of $2.90 \times 10^{-5} \Msun$. AGB, more specifically TP-AGB, nucleosynthesis and evolution are not calibrated in the $6.5-8.0\Msun$ mass range. The peak \iso{26}Al production from the uncalibrated masses are an order of magnitude higher than the \emph{Monash16} models and dominate the \iso{26}Al stellar yields from our AGB stars.

Our modifications to \textsc{binary\_c} aim to reduce the impact from the uncalibrated mass range by reducing minimising the mass range it influences (see Section \ref{sec:PopSynth} for discussion). The production of \iso{26}Al is very sensitive to HBB temperatures and the slight increase in $T_{\rm BCE, max}$, triggered by increased core mass as the stellar structure of models transition to fits to \emph{Pols98}, leads to the increased \iso{26}Al production compared to our standard \textsc{binary\_c} models. 

Our modification to \textsc{binary\_c} reduces the mass range of uncalibrated AGB evolution from $6.5-8.0\Msun$ to $8.0-8.3\Msun$. Fig. \ref{fig:Al26Single} shows there is still a spike in the uncalibrated mass range ($8.0-8.3\Msun$) in the \iso{26}Al yields, peaking at $8.29\Msun$ with $7.86 \times 10^{-5} \Msun$. The underlying reason for this spike, in both the standard and modified versions of \textsc{binary\_c}, is the transition to \emph{Pols98} fits leading to over-massive cores and increased HBB temperatures. The degree of this increase is reduced in our modified \textsc{binary\_c} due to the relatively small $0.3 \Msun$ uncalibrated region before stars are massive enough to be treated as massive stars. We consider the yields from our modified version of \textsc{binary\_c} to be more reasonable than the standard \textsc{binary\_c} as the mass of \iso{26}Al produced by the $6.5-8.0\Msun$ stars more closely follow the trends of the \emph{Monash16} and \citet{Doherty2014} (VW-93 case, see Fig. \ref{fig:Al26Single}) models and the peak \iso{26}Al production in the uncalibrated mass ranges are reduced by 48\%. The \iso{26}Al yields presented by \citet{Doherty2014}, which covers a mass range of $6.5\Msun < M_0 < 9.0\Msun$ at solar metallicity, do not exhibit any spike in \iso{26}Al similar to the behaviour shown in Fig. \ref{fig:Al26Single} for the standard \citet{Vassiliadis1993} AGB mass loss case.

The modified \citet{Vassiliadis1993} AGB mass loss (VW-M) case case from \citet{Doherty2014} presented in Fig. \ref{fig:Al26Single} has an \iso{26}Al yield peak at $7.5\Msun$ similar to the \iso{26}Al yields from the standard version of \textsc{binary\_c}. Since there are no models for the VW-M case with masses $< 7\Msun$ from \citet{Doherty2014}, we cannot determine if this is a spike in the \iso{26}Al yields similar to the standard version of \textsc{binary\_c}, or if the \iso{26}Al yields would gradually increase with mass like the \emph{Monash16} stars as shown in Fig. \ref{fig:Al26Single}. 

\subsubsection{Weighted population yields}
\label{sec:weightedYields}

\begin{figure*}
\begin{subfigure}{\textwidth}
  \centering
  \includegraphics[width=.8\linewidth]{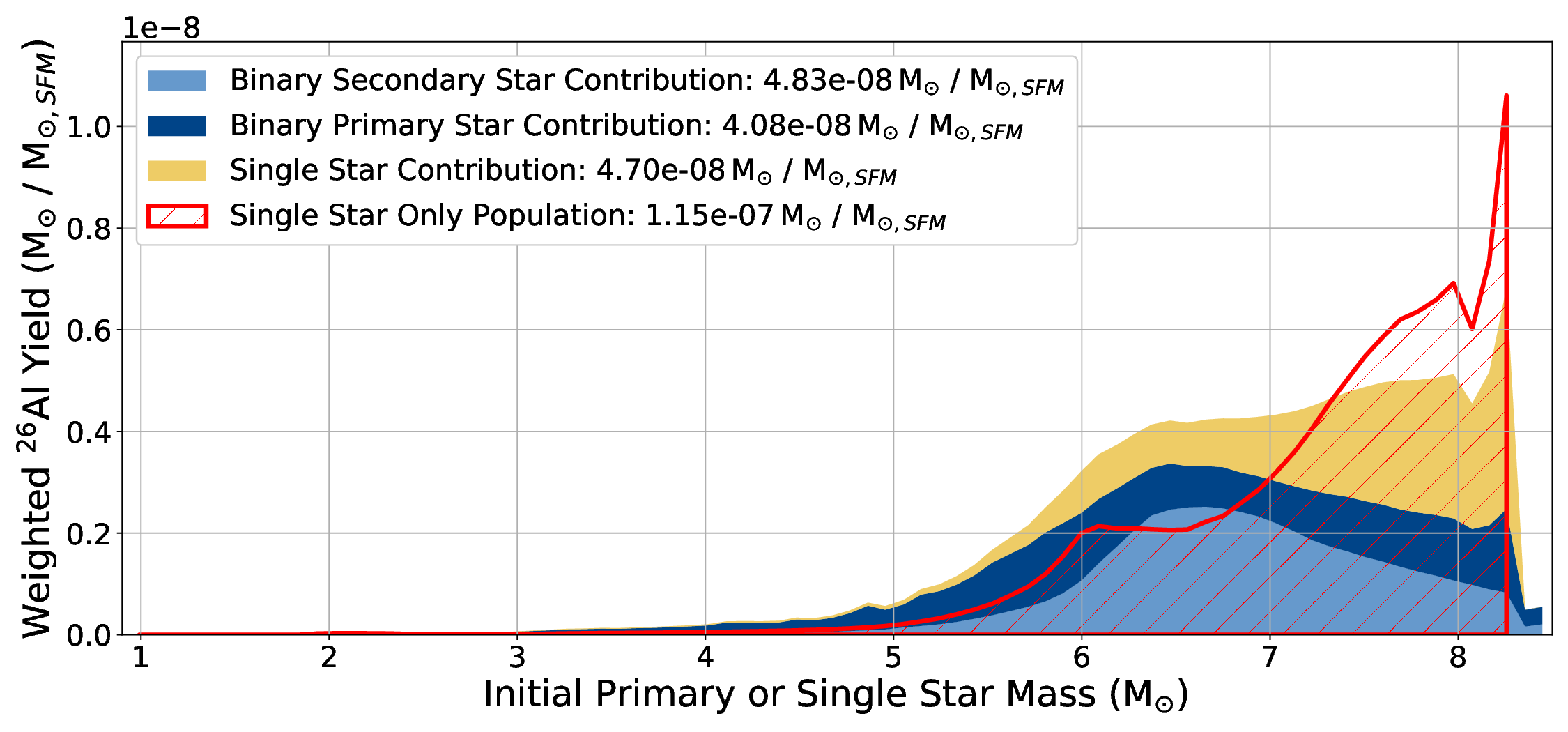}  
\end{subfigure}
\begin{subfigure}{\textwidth}
  \centering
  \includegraphics[width=.8\linewidth]{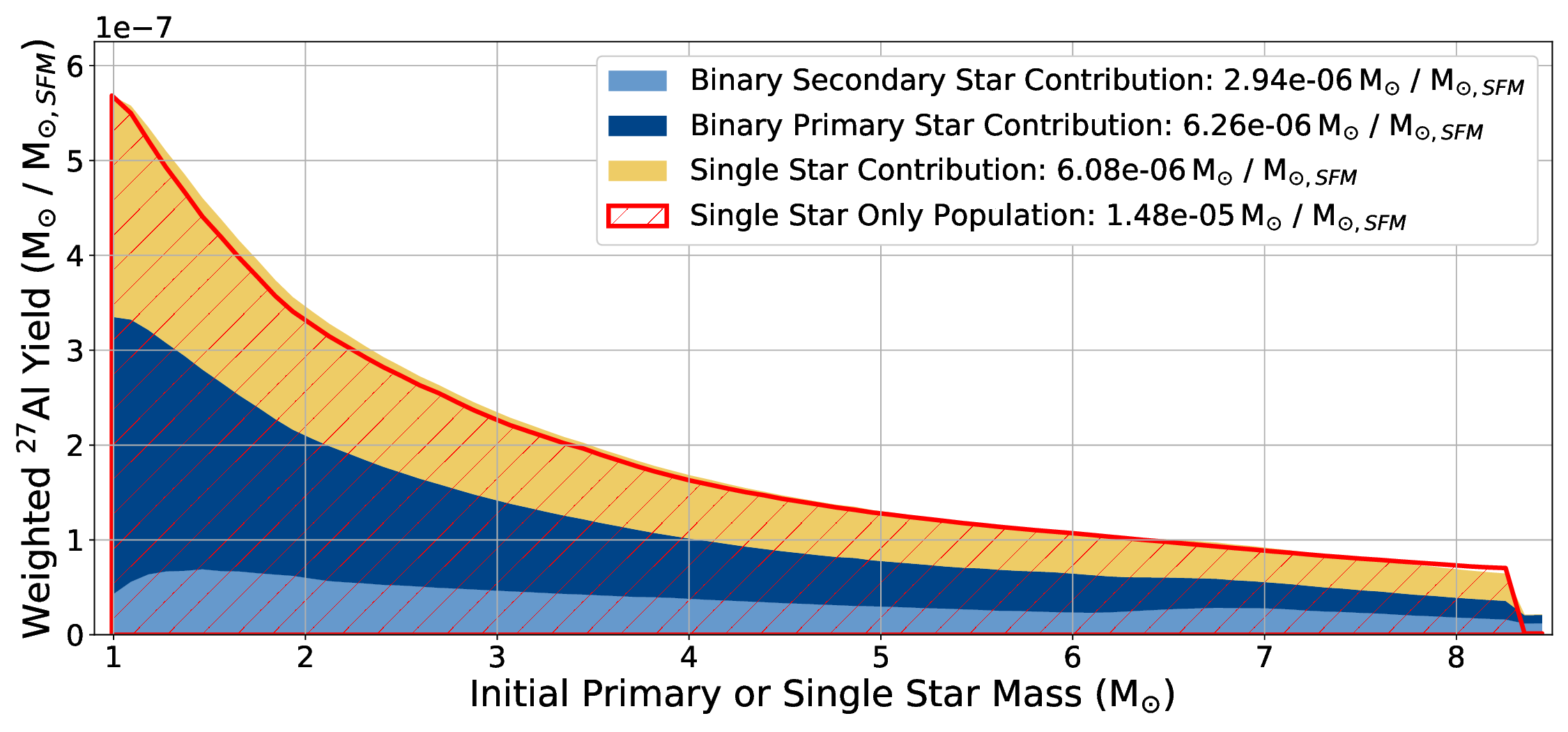}  
\end{subfigure}

\caption{Stacked plots showing the weighted population yields of $^{26}$Al (top) and \iso{27}Al (bottom) in solar mass per solar masses of star-forming material ($\Msun/{\rm M_{\odot,SFM}}$) vs the initial single or binary primary-star mass. The population has a binary fraction of 0.5. The contributions to total population yield from the single, binary primary, and binary secondary stars are shown. The single-star portion of the population is yellow, the binary primary is dark blue, and the secondary star portion is light blue. The yield produced by a population of single-stars is shown by the red hatched area. The weighted yields are computed using an $80 \times 80 \times 80$ grid in \textsc{binary\_c}. The top panel shows that the inclusion of binaries results in 18\% more \iso{26}Al being produced. The binary primary stars produce the majority of \iso{26}Al at masses $<6.00\Msun$ and the secondary stars at $6.00-7.23\Msun$. The bottom panel shows that the inclusion binary systems into our population alters the population \iso{27}Al yield by $<5\%$.}
\label{fig:PopYields}
\end{figure*}

The top panel of Fig. \ref{fig:PopYields} shows the weighted \iso{26}Al population yield in units of solar masses per solar mass of star-forming material ($\Msun/{\rm M_{\odot,SFM}}$) from a population with a binary fraction of 0.5, plotted against the initial  primary-star mass. The bottom panel of Fig. \ref{fig:PopYields} shows the equivalent figure for \iso{27}Al. 

Table \ref{tab:PopYieldFrac} shows the total weighted population yield of \iso{26}Al and \iso{27}Al. We consider the influence of binary stellar evolution on \iso{27}Al, since the ratio of \iso{26}Al/\iso{27}Al is commonly used when considering the production of \iso{26}Al (e.g. in dust grains). 

The top panel of Fig. \ref{fig:PopYields} shows that binary influence leads to a net overproduction of \iso{26}Al in systems with initial primary masses $ \lesssim 7.6\Msun$ and some underproduction in systems where $M_{\rm{1,0}} \gtrsim 7.6\Msun$, when compared to a single-star population. Table \ref{tab:PopYieldFrac} shows that when we have a binary fraction of 0.5 we see an overall \iso{26}Al yield increase of $18\%$ relative to our single-star population. If we increase the binary fraction to 0.75, Table \ref{tab:PopYieldFrac} shows a $25\%$ increase in the \iso{26}Al weighted population yield. The binary evolutionary mechanisms influencing these changes in \iso{26}Al production are explored in Sections \ref{sec:CloserLook} and \ref{sec:IMS_Results}.

The bottom panel of Fig. \ref{fig:PopYields} and Table \ref{tab:PopYieldFrac} shows that binary evolution has little influence over \iso{27}Al yields. The total weighted population yields of all populations, including binaries, remain within 5\% of the total \iso{27}Al ejected by our single-star population. This allows us to attribute any change from single-star \iso{26}Al/\iso{27}Al ratios to \iso{26}Al.

\begin{table*}
	\centering
	\caption{\iso{26}Al and \iso{27}Al weighted population yields from our populations of binary fraction 0.0, 0.25, 0.50, 0.75, and 1.0, and their ratios to our population of only single-stars.}
	\label{tab:PopYieldFrac}
	\begin{tabular}{l|c|ccccc} 
		\hline
		Isotope & & \multicolumn{5}{c}{Binary Fraction}\\
        & & 0.00 & 0.25 & 0.5 & 0.75 & 1.0 \\
		\hline
        \multirow{2}{*}{\iso{26}Al}
            & Weighted population yield ($\Msun/{\rm M_{\odot,SFM}}$) & $1.147 \times 10^{-7}$ & $1.266 \times 10^{-7}$ & $1.361 \times 10^{-7}$ & $1.431 \times 10^{-7}$ & $1.499 \times 10^{-7}$ \\
		      & Ratio pop. yield incl binaries / single only & 1.000 & 1.104 & 1.187 & 1.248 & 1.306 \\
        \hline
        \multirow{2}{*}{\iso{27}Al}
            & Weighted population yield ($\Msun/{\rm M_{\odot,SFM}}$) & $1.483 \times 10^{-5}$ & $1.507 \times 10^{-5}$ & $1.527 \times 10^{-5}$ & $1.532 \times 10^{-5}$ & $1.545 \times 10^{-5}$ \\
		      & Ratio pop. yield incl binaries / single only & 1.000 & 1.018 & 1.023 & 1.034 & 1.042 \\
		\hline
	\end{tabular}
\end{table*}

\subsubsection{A closer look at individual binary systems}
\label{sec:CloserLook}
\begin{figure*}
    \centering
    \begin{subfigure}[b]{0.475\textwidth}
        \centering
        \includegraphics[width=\textwidth]{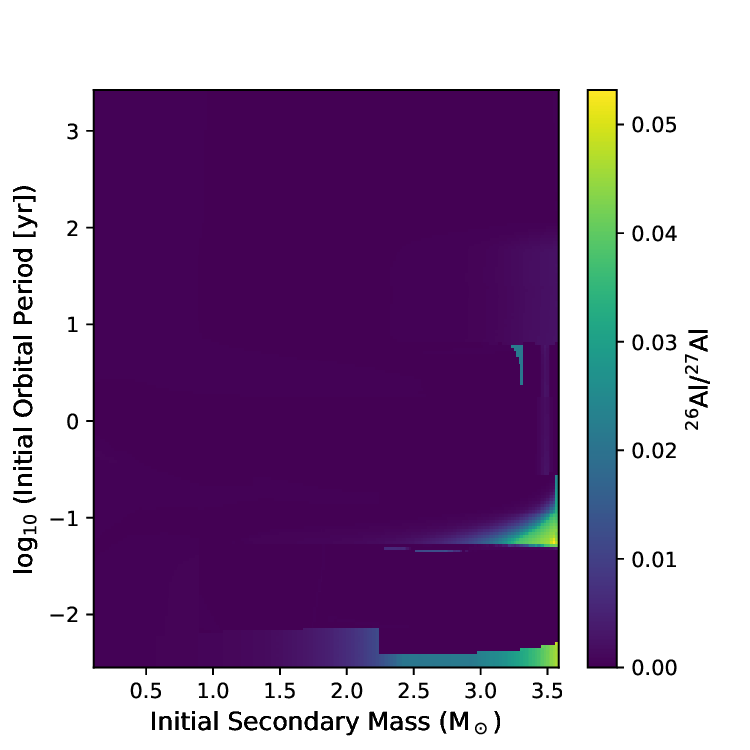}
        \caption[]{\iso{26}Al/\iso{27}Al presented with a linear colour scale}%
        \label{fig:BinRatios3a}
        {}    
    \end{subfigure}
    \hfill
    \begin{subfigure}[b]{0.475\textwidth}
        \centering 
        \includegraphics[width=\textwidth]{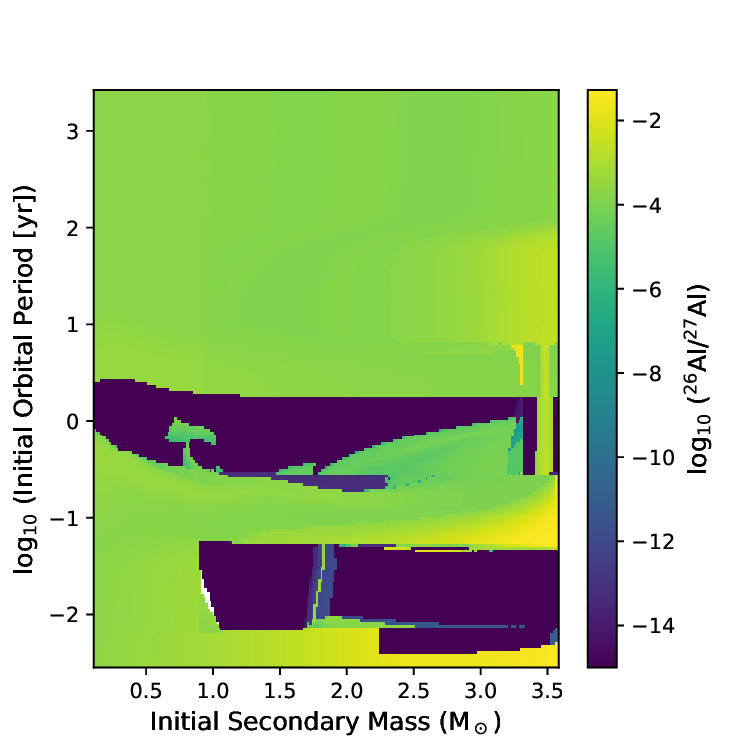}
        \caption[]{\iso{26}Al/\iso{27}Al presented with a log colour scale}%
        \label{fig:BinRatios3b}
        {}    
    \end{subfigure}
    \caption[] {\iso{26}Al/\iso{27}Al calculated from the stellar yields of binary systems with initial primary mass of $3.59\Msun$. Panel (a) presents the data using a linear scale for the colouring to better highlight \iso{26}Al overproduction where mergers allow for stars to gain sufficient mass for HBB. Panel (b) shows the same data as panel (a) but using a logarithmic colour scale to highlight \iso{26}Al underproduction where binary evolution either shortens or prevents stars from entering the TP-AGB. Areas of white-space in (b) indicate \iso{26}Al/\iso{27}Al = 0. Data are generated using a $200 \times 200$ grid in our modified version of \textsc{binary\_c}. The ejected material from the $3.59\Msun$ single-star has an \iso{26}Al/\iso{27}Al ratio of $2.1\times10^{-4}$ (${\rm log_{10}}\left[\text{\iso{26}Al/\iso{27}Al}\right]=-3.7$).} 
    \label{fig:BinRatios3}
\end{figure*}

\begin{figure*}
    \centering
    \begin{subfigure}[b]{0.475\textwidth}
        \centering
        \includegraphics[width=\textwidth]{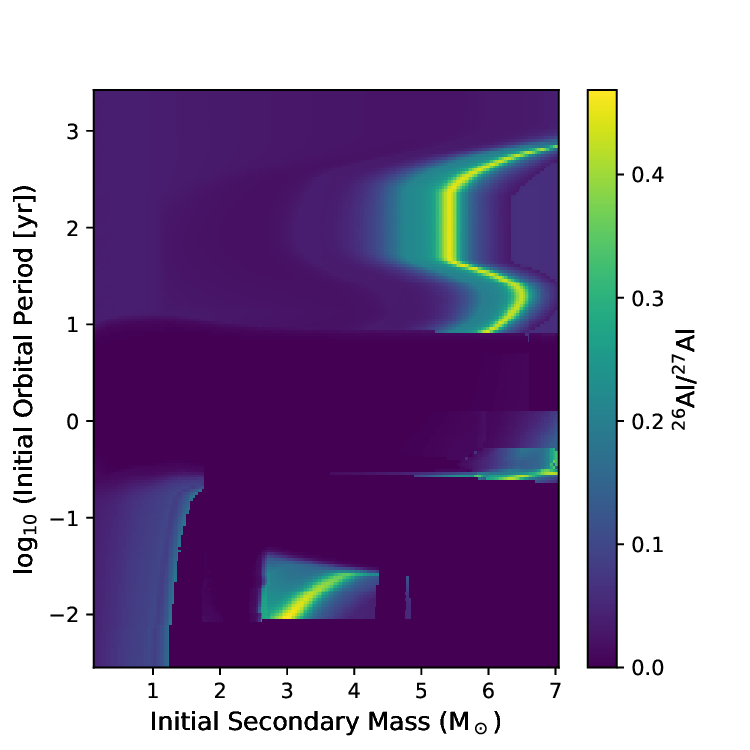}
        \caption[]{\iso{26}Al/\iso{27}Al presented with a linear colour scale}%
        \label{fig:BinRatios7a}
        {}    
    \end{subfigure}
    \hfill
    \begin{subfigure}[b]{0.475\textwidth}
        \centering 
        \includegraphics[width=\textwidth]{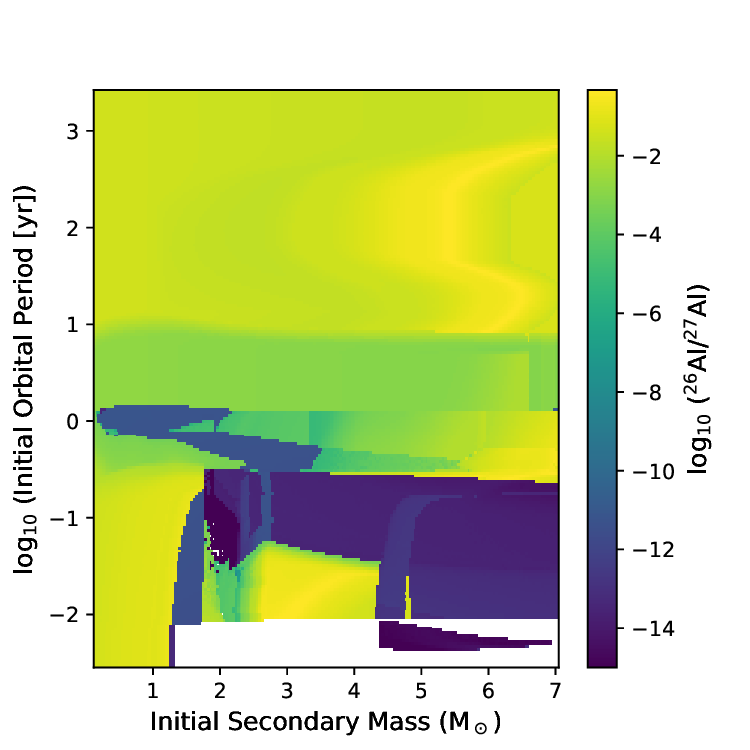}
        \caption[]{\iso{26}Al/\iso{27}Al presented with a log colour scale}%
        \label{fig:BinRatios7b}
        {}    
    \end{subfigure}
    \caption[] {As Fig. \ref{fig:BinRatios3} for an initial primary mass of $7.06\Msun$. The single-star $7.06\Msun$ case has an \iso{26}Al/\iso{27}Al ratio from the ejected material of $3.9\times 10^{-2}$. Mergers can result in massive stars which explode without entering the TP-AGB. Mergers and WRLOF may also lead to \iso{26}Al overproduction as discussed in Section \ref{sec:IMS_Results}.} 
    \label{fig:BinRatios7}
\end{figure*}

To understand which stellar conditions correspond to \iso{26}Al production in stellar populations, we focus here on individual binary systems. Figures \ref{fig:BinRatios3} and \ref{fig:BinRatios7} present the \iso{26}Al/\iso{27}Al ratios calculated from the stellar yields of various binary systems with primary masses of $3.59\Msun$ and $7.07\Msun$ respectively. 

Fig. \ref{fig:BinRatios3a} shows there are some binary system with \iso{26}Al/\iso{27}Al ratios over two orders of magnitude greater than that of a single $3.59\Msun$ star (\iso{26}Al/\iso{27}Al $ = 2.1\times10^{-4}$). The \iso{26}Al overproduction in stars with initial masses $\lesssim 5\Msun$ is primarily due to stellar mergers, which produce new stars of sufficient mass to undergo HBB. More precisely, the overproducing stars with initial secondary mass $M_{\rm {2,0}} \gtrsim 3\Msun$ and initial orbital period $p_{\rm 0} \sim 0.1$ yr in Fig. \ref{fig:BinRatios3a} merge after the primary ascends the GB and the secondary is on the MS. The overproducing stars with $p_{\rm 0} < 0.01$ yr in Fig. \ref{fig:BinRatios3a} merge when both stars are on the MS, resulting in a new star which effectively evolves as a single-star with sufficient mass for HBB. 

Because low- and intermediate-mass stars primarily produce, transport, and eject \iso{26}Al during the TP-AGB phase, the regions of underproduction (log[\iso{26}Al/\iso{27}Al] $\lesssim -10$) shown in Fig. \ref{fig:BinRatios3b} reflect binary systems that do not experience or have a short TP-AGB phase as a result of binary interactions. The systems with initial orbital periods $p_{\rm 0} < 0.1$ yr interact when the primary-star is on the GB, and the systems with initial orbital periods $p_{\rm 0} \sim 1$ yr mostly interact while the primary is on the AGB branch. Unstable mass transfer and CE events lead to these stars being stripped of their envelopes and truncating, or not entering, the TP-AGB phase.

The key difference between a single $3.59\Msun$ and a $7.06\Msun$ star, in the context of \iso{26}Al production, is that a single $7.06\Msun$ star has sufficient mass to synthesize \iso{26}Al through HBB. Fig. \ref{fig:BinRatios7a} shows there are four distinct regions of \iso{26}Al overproduction. The systems with initial secondary star mass $\lesssim 1.5\Msun$ and initial orbital period $\lesssim 0.1 \, \mathrm{yr}$ merge while both stars are either on the MS, or when the primary is crossing the HG and the secondary is on the MS. The systems with secondary masses $\sim 3-4\Msun$ and initial orbital period $\sim 0.01 \, \mathrm{yr}$ and the systems with secondary masses $\gtrsim 6\Msun$ and orbital periods $\sim 1 \, \mathrm{yr}$ in Fig. \ref{fig:BinRatios7a} commonly experience multiple RLOF and CE events before they merge. The stellar types at the time of the merger are often a CO-WD, with either a naked-He or HG companion, or a merger between a naked-He and HG star. Finally, in Fig. \ref{fig:BinRatios7} we see \iso{26}Al overproduction in systems with initial orbital periods $\gtrsim 10 \, \mathrm{yr}$. This \iso{26}Al originates from the secondary stars which gain, in extreme cases, $\sim 3\Msun$ via WRLOF.

Similarly to the $3.59\Msun$ primary stars presented in Fig. \ref{fig:BinRatios3b}, the distinct regions of \iso{26}Al underproduction visible in our $7.06\Msun$ stars in Fig. \ref{fig:BinRatios7b} are attributed to binary influence preventing the stars from experiencing a complete TP-AGB phase. Also similarly to the systems presented in Fig. \ref{fig:BinRatios3}, some underproduction is explained by binary interactions with a GB or AGB star. Fig. \ref{fig:BinRatios7b} also shows a region of zero \iso{26}Al yield in some systems with an initial orbital period shorter then $0.01 \, \mathrm{yr}$. The systems producing zero \iso{26}Al experience mergers when both stars are on the MS, resulting in a single-star with mass $\gtrsim 8.30\Msun$. In our modified \textsc{binary\_c}, stars of mass $\gtrsim 8.30\Msun$ are considered massive stars and explode without ejecting any \iso{26}Al via stellar winds. In this study we are only considering \iso{26}Al released via stellar winds and do not consider \iso{26}Al synthesized within supernovae.

\begin{figure*}
	\includegraphics[width=\textwidth]{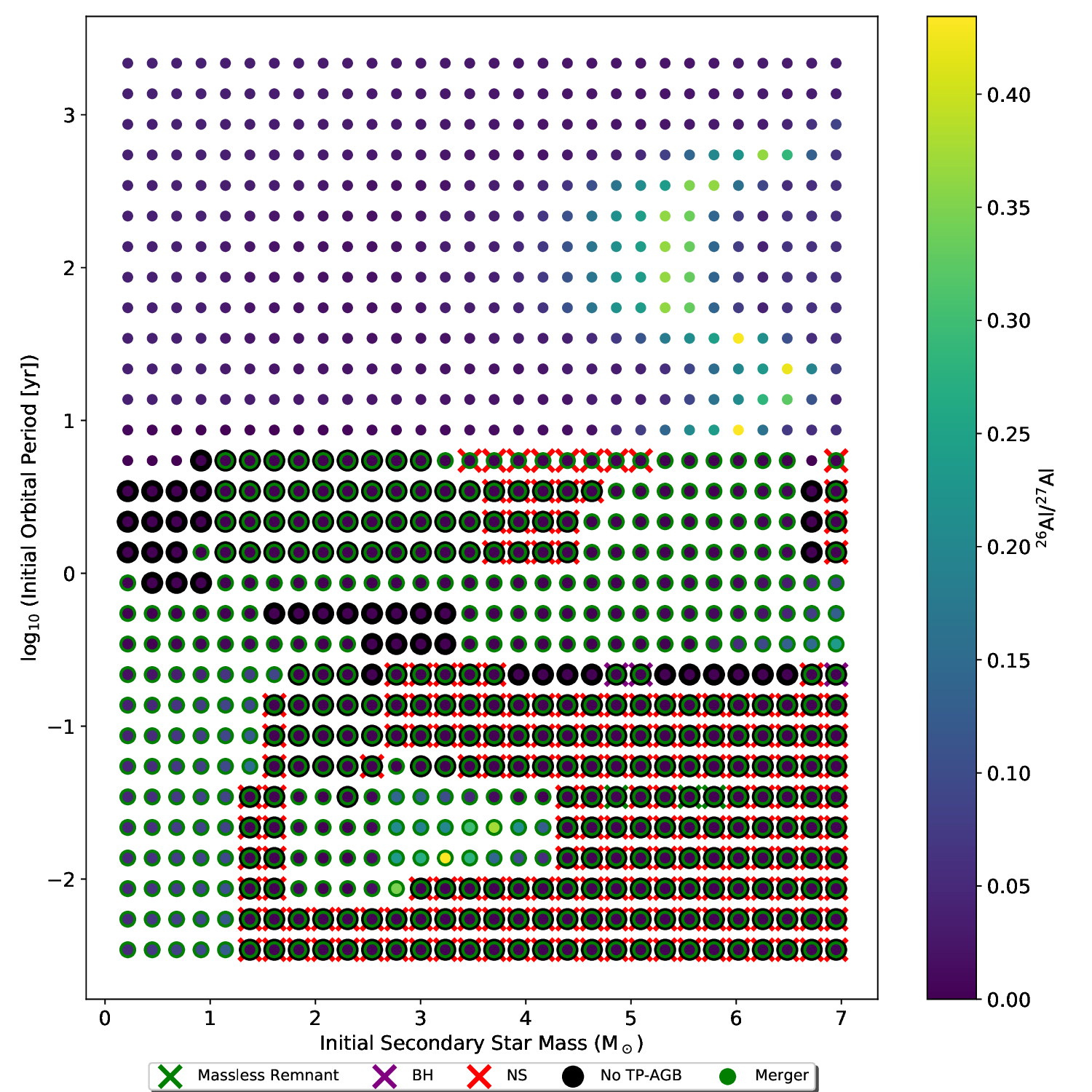}
    \caption{Similar to Fig. \ref{fig:BinRatios7a} but highlighting the stellar evolution of the primary (or post-merger) stars. Created using a $30 \times 30$ grid in our modified version of \textsc{binary\_c}. A cross indicates the primary-star explodes during the life of the simulation. A green cross indicates a Type 1.5 supernova \citep{Lau2008}, a purple cross indicates an explosion forming a BH, and a red cross indicates an explosion resulting in the formation of a NS. The presence of a black ring indicates that the binary primary-star does not experience the TP-AGB phase, and a green ring indicates that the binary system merges during the life of the simulation. We find supernovae, which leave NS remnants, often result in \iso{26}Al underproduction. Systems with an orbital period $\gtrsim 10 \, \mathrm{yr}$ neither merge nor explode, yet they overproduce \iso{26}Al. This is the result of efficient WRLOF onto the secondary star.}
    \label{fig:Evol7}
\end{figure*}

Fig. \ref{fig:Evol7} shows the \iso{26}Al/\iso{27}Al ratios calculated from the stellar yields of binaries with primary masses equal to $7.06\Msun$ and highlights the stellar evolution of the binary's primary and merged star (if applicable). Fig. \ref{fig:Evol7} shows us the evolutionary phenomena affecting \iso{26}Al production. For example, the TP-AGB phase is very important for \iso{26}Al production and all stars not experiencing the TP-AGB underproduce \iso{26}Al compared to single-stars of identical mass. These stars either all have their envelopes stripped, or they merge and later explode. We also see in Fig. \ref{fig:Evol7} that the overproducing systems with initial orbital periods $\lesssim 1 \, \mathrm{yr}$ all merge. Fig. \ref{fig:Evol7} also shows that none of the systems with $p_{\rm 0} \gtrsim 10 \, \mathrm{yr}$ explode or merge, yet they still overproduce \iso{26}Al compared to our single $7.06\Msun$ star. These systems all experience efficient WRLOF, and the majority \iso{26}Al from these systems are ejected by the secondary stars.

\subsubsection{Binary systems with initial primary mass $> 5\Msun$}
\label{sec:IMS_Results}

\begin{figure}
	\includegraphics[width=\columnwidth]{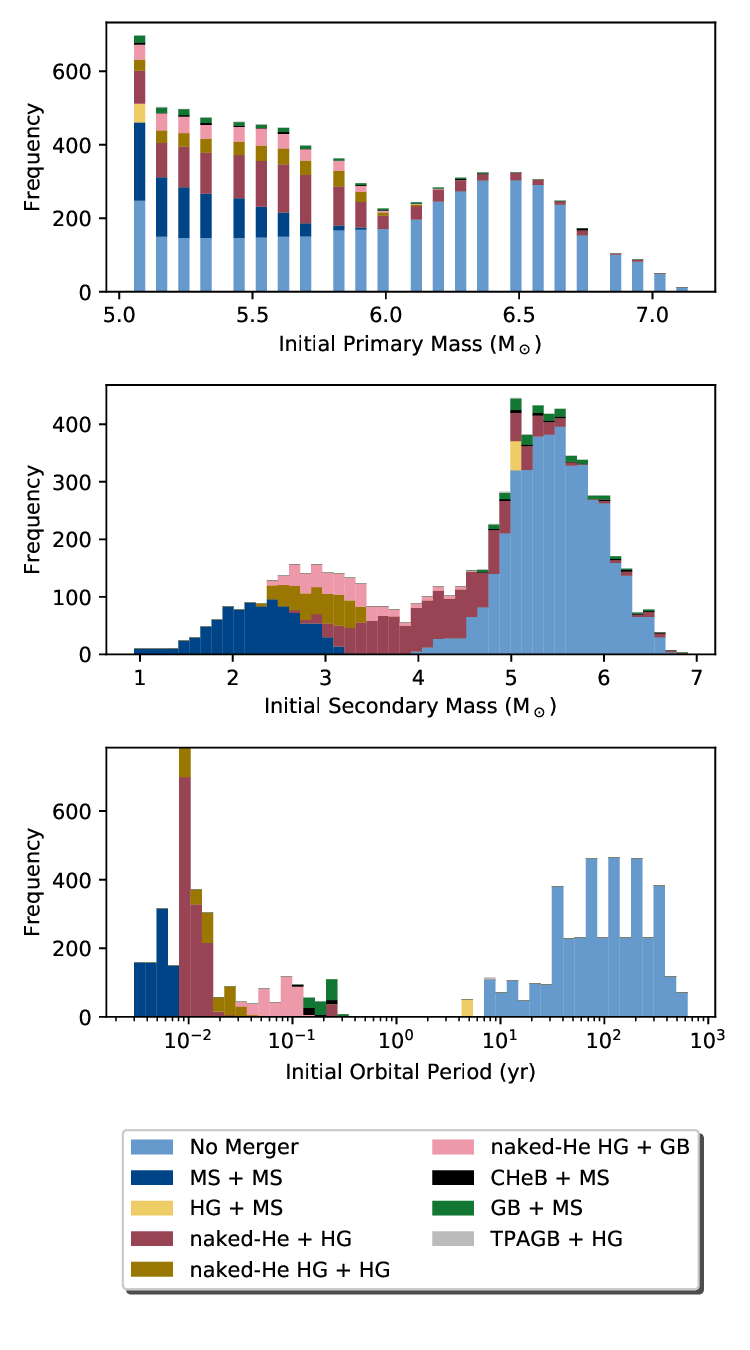}
    \caption{Initial conditions of binary systems with \iso{26}Al/\iso{27}Al at least ten times that of a single-star of the same initial mass calculated from the stellar yields. In this figure we consider primary initial stellar mass $> 5\Msun$. Coloured according to the stellar types/evolutionary phase of the binary at the time of merger. Binary data is extracted from $80 \times 80 \times 80$ \textsc{binary\_c} grid. A total of 7297 systems from this grid were identified as \iso{26}Al overproducers. The "No Merger" channel is the most common evolutionary channel with 4047 systems identified are often wider binaries with initial orbital periods of $10-10^3 \, \mathrm{yr}$ and with initial secondary mass $\sim 4.5-7\Msun$.}
    \label{fig:OverAchieve}
\end{figure}

The top panel of Fig. \ref{fig:PopYields} shows that binary systems of initial primary mass $\gtrsim 5\Msun$ produce the majority of \iso{26}Al in a low- and intermediate-mass population, therefore our further discussion focuses on this mass range. Fig. \ref{fig:OverAchieve} presents the initial conditions of binary systems which achieve an \iso{26}Al/\iso{27}Al abundance ratio of at least 10 times that of a single-star of identical mass to the primary-star. Fig. \ref{fig:OverAchieve} shows the most common ($55\%$ of systems identified in Fig. \ref{fig:OverAchieve}) evolutionary channel for \iso{26}Al overproduction is the "No Merger" channel. The "No Merger" channel represents binary systems with efficient WRLOF allowing \iso{26}Al overproduction by the secondary stars. All other evolutionary channels highlighted in Fig. \ref{fig:OverAchieve} show that mergers at various evolutionary stages also lead to \iso{26}Al overproduction. The most common merger case ($18\%$ of systems identified in Fig. \ref{fig:OverAchieve}) leading to \iso{26}Al overproduction is the "naked-He + HG" case.

\begin{figure}
    \centering
        \centering
        \includegraphics[width=\columnwidth]{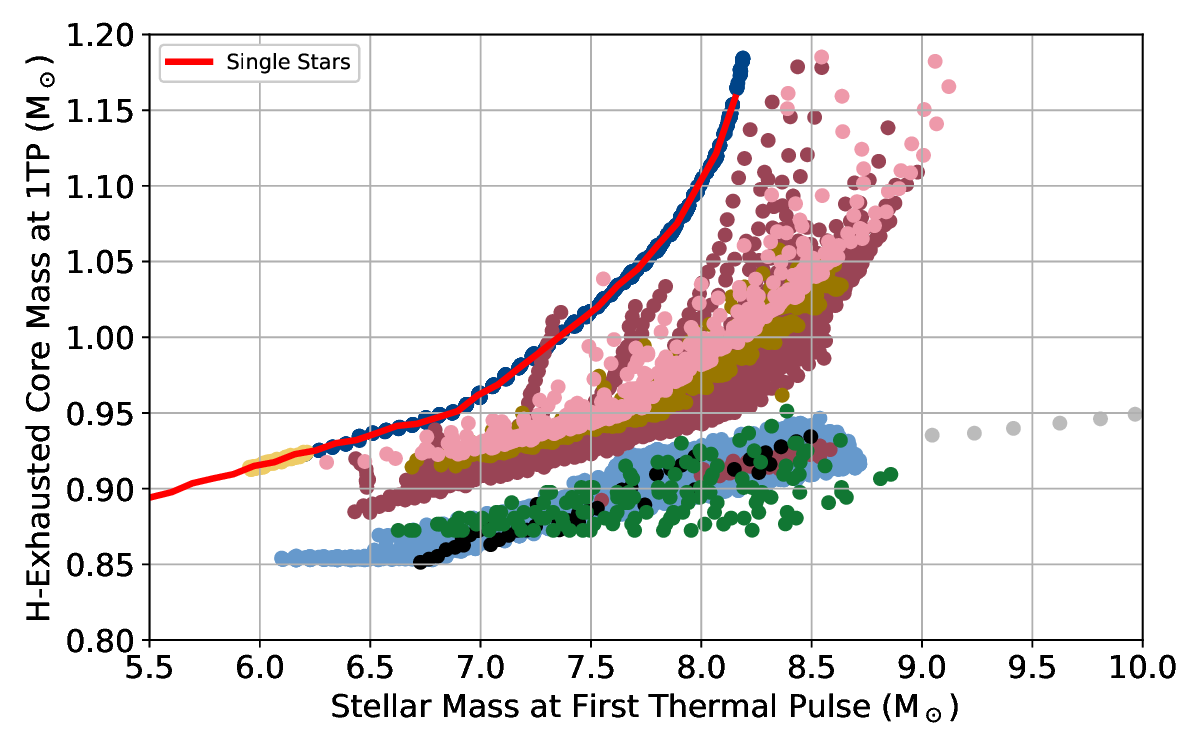}
        \caption[]{Core mass at first thermal pulse vs. mass at first thermal pulse in all systems identified in Fig. \ref{fig:OverAchieve}. See Fig. \ref{fig:OverAchieve} for marker colour legend. If the system merges this plot shows data from the post-merger object, and if overproduction is the result of WRLOF (the "No Merger" case) the plot shows data from the secondary star (the overproducing star in the system). Data from single-star models are shown as the red line and terminate at $M \approx 8.30\Msun$ as more-massive single-stars explode and do not experience the TP-AGB. We find most of our \iso{26}Al overproducing stars either have core mass equal to that of a single-star model of identical mass or less.}%
        \label{fig:1TPMc}    
\end{figure}

\begin{figure} 
        \centering 
        \includegraphics[width=\columnwidth]{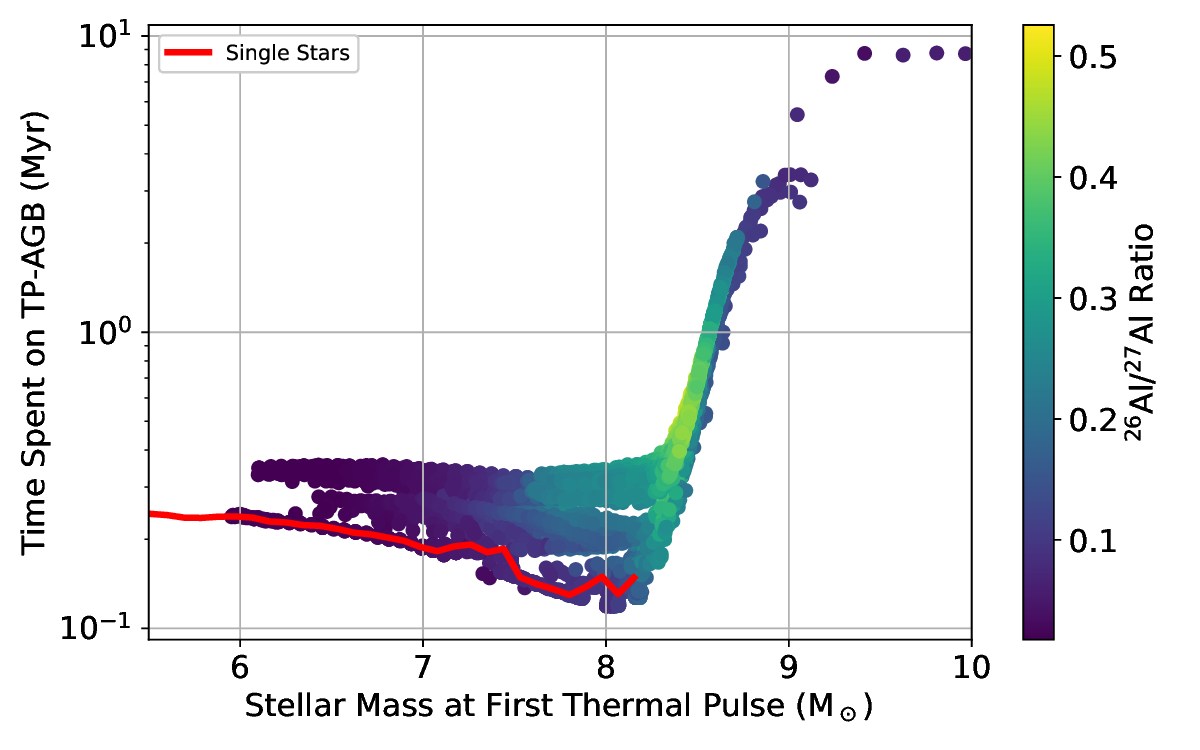}
        \caption[]{Time spent on the TP-AGB vs mass at first thermal pulse for all systems identified in Fig. \ref{fig:OverAchieve}. Marker colours show the \iso{26}Al/\iso{27}Al abundance ratio calculated from the stellar yields. If the system merges this plot shows data from the post-merger object, and if overproduction is the result of WRLOF the plot shows shows data from the secondary star, which is the overproducer. The single-stars are the red line. We find most of our \iso{26}Al overproducing stars experience relatively long TP-AGB phases compared to single-stars of identical mass.}%
    \label{fig:TpTime}
\end{figure}

The stellar structures of the \iso{26}Al overproducing stars identified in Fig. \ref{fig:OverAchieve} are now explored in more detail. Fig. \ref{fig:1TPMc} shows the core mass vs. the total mass of the overproducing \iso{26}Al stars identified in Fig. \ref{fig:OverAchieve} at the first thermal pulse of the TP-AGB phase. We find that due to binary evolution, through either a merger or stable mass accretion, most systems which overproduce \iso{26}Al almost enter the TP-AGB phase with a relatively low-mass core for their total mass when compared to single-stars. This is due to the standard and modified versions of \textsc{binary\_c} using the parameter $M_{\rm PostMS}$ in the fitting formulae for the core masses of evolved stars (see Section \ref{sub:CoreMassEAGB}). 

There are four systems identified in Fig. \ref{fig:1TPMc} that enter the TP-AGB with relatively massive cores for their total mass, with their cores up to $0.02\Msun$ more massive than single-stars of equivalent TP-AGB mass. These systems have initial primary-star masses of $5.2-5.4\Msun$ and enter the TP-AGB with masses $7.3-7.6\Msun$. Although these systems produce less \iso{26}Al than our single $7.3-7.6\Msun$ stars, they produce at least an order of magnitude more \iso{26}Al than our single $5.2-5.4\Msun$ stars. 

Fig. \ref{fig:TpTime} shows the time the stars identified as \iso{26}Al overproducing in Fig. \ref{fig:OverAchieve} spend in the TP-AGB phase vs. the total mass of the overproducing star at the first thermal pulse. Fig. \ref{fig:TpTime} shows that most systems which overproduce \iso{26}Al spend more time on the TP-AGB phase than single-stars of the same total mass. Fig. \ref{fig:TpTime} also shows that stars entering the TP-AGB phase with a total mass of $\sim 8.4\Msun$ have the highest \iso{26}Al/\iso{27}Al. However it is important to remember that the TP-AGB evolution physics in \textsc{binary\_c} is calibrated up to $8\Msun$ using approximations based on single-star evolution, so we cannot definitively conclude that our synthetic models obey the stellar evolution equations when mass is greater than $8\Msun$. Further investigation with a detailed model is required to verify this result.

\begin{figure}
	\includegraphics[width=\columnwidth]{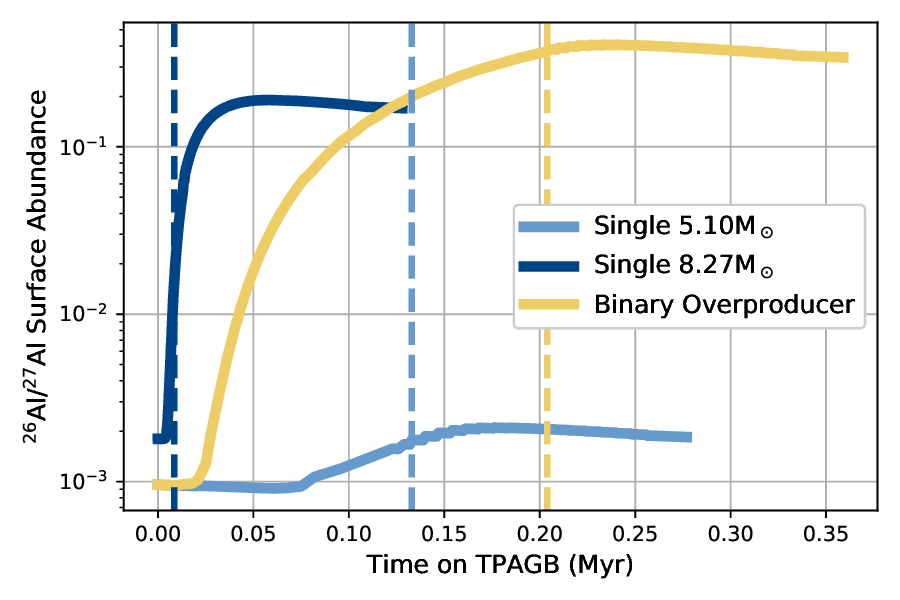}
    \caption[]{The surface abundance ratio of \iso{26}Al/\iso{27}Al vs time on the TP-AGB in an overproducing binary star produced with our modified version of \textsc{binary\_c}. The initial conditions of the binary system are: $M_{\rm {1,0}} = 5.10\Msun$, $M_{\rm {2,0}} = 5.05\Msun$, and $p_{\rm 0} = 55$ days. The stars merge when the primary enters the GB and the secondary is on the MS. The post-merger star enters the TP-AGB with a mass of $8.26\Msun$ and a core mass of $0.87\Msun$. For comparison, the surface \iso{26}Al mass fraction for a single $5.10\Msun$ and a single $8.27\Msun$ star are included. The dashed lines indicate the onset of the superwind which triggers a period of rapid mass loss. We find the \iso{26}Al/\iso{27}Al surface abundance ratio of our post-merger overproducer peaks at 0.41, 2.16 times the peak surface abundance ratio of the single $8.27\Msun$ and 205 times higher than our single $5.10\Msun$ star. This overproducer has an \iso{26}Al stellar yield of $2.00 \times 10^{-4} \Msun$.}
    \label{fig:SurfaceAl26}
\end{figure}

Fig. \ref{fig:SurfaceAl26} shows the surface \iso{26}Al/\iso{27}Al abundance ratio of an overproducing star with $M_{\rm {1,0}} = 5.10 \Msun$, $M_{\rm {2,0}} = 5.05 \Msun$, and $p_{\rm 0} = 0.15$ yr post-merger throughout the duration of the TP-AGB upon which it enters with a mass of $8.26 \Msun$ and a core mass of $0.87 \Msun$. Fig. \ref{fig:SurfaceAl26} also shows the surface abundance ratio of a single $5.10\Msun$ star which shares the same initial mass as the overproducing primary-star, and of a single $8.27\Msun$ star which enters the TP-AGB with a total mass of $8.19\Msun$ and a core mass of $1.18\Msun$. Our overproducing binary star experiences 72 thermal pulses and has a peak \iso{26}Al/\iso{27}Al surface ratio of 0.41, 2.16 times that of our single $8.27\Msun$ star (peak surface \iso{26}Al/\iso{27}Al = 0.19) and higher than that of our single $5.10\Msun$ star (\iso{26}Al/\iso{27}Al = $2.10 \times 10^{-3}$). The superwind \citep[see][]{Vassiliadis1993} in our binary overproducer begins after it spends $0.204 \, \mathrm{Myr}$ on the TP-AGB, compared to $8,000 \, \mathrm{Myr}$ in our single $8.27\Msun$ and $0.133 \, \mathrm{Myr}$ in our single $5.10\Msun$ stars. Therefore, our binary overproducer spends about about 25 times more time HBB than our single $8.27\Msun$ star, and 1.5 times longer than our single $5.10\Msun$ star. This binary overproducer has an \iso{26}Al yield of $2.00 \times 10^{-4} \Msun$ and an \iso{26}Al/\iso{27}Al ratio of $0.31$ as calculated from the ejected mass. The \iso{26}Al/\iso{27}Al ratio calculated from the stellar yield is always lower than the peak surface \iso{26}Al/\iso{27}Al abundance ratio as the stellar yield is calculated using dynamic surface abundances over the lifetime of the star (Eq. \ref{eq:SinYield}). The reduced mass loss rate and delayed onset of the superwind is attributed to \iso{26}Al overproducers having a reduced surface luminosity and smaller radii compared to single-stars of the same total mass due to their less massive cores.

\subsection{Monash models}
\label{sub:Monash}

To test the results of our modified version of \textsc{binary\_c}, we look at the stellar structure and surface abundances from our $5+2\Msun$ and $5+3\Msun$ Monash stellar-models at solar metallicity, $Z=0.014$. We present an overview of our results in Table \ref{tab:MonashModels}.

\begin{table*}
	\centering
	\caption{Summary of the results from our Monash stellar-models. We include the core mass at the tip of the red giant-branch $M_{\rm c, RGBtip}$, H-exhausted core mass after core He depletion $M_{\rm c, PostCHeB}$, H-exhausted core mass at the first thermal pulse $M_{\rm c,1tp}$, the highest temperature reached in the bottom of the convective envelope $T_{\rm bce}^{\rm max}$, lifetime of the CHeB phase $\tau_{\rm CHeB}$, lifetime of the TP-AGB $\tau_{\rm TPAGB}$, mass lost during the final interpulse period before model termination $M_{\rm lost} ^{\rm final}$, the \iso{26}Al stellar yield $y_{\rm 26Al}$, and the \iso{26}Al/\iso{27}Al ratio calculated from the stellar yields. We do not calculate the \iso{26}Al yield from our $5+2\Msun$ star and we only consider results related to the TP-AGB for our $5+3\Msun$ star.}
	\label{tab:MonashModels}
	\begin{tabular}{l|c|cccccccc} 
		\hline
        \multicolumn{10}{c}{Single-star Monash models} \\
        Model & $M_{\rm c, RGBtip}$ ($\Msun$) & $M_{\rm c, PostCHeB}$ ($\Msun$) & $M_{\rm c,1TP}$ ($\Msun$) & $T_{\rm bce}^{\rm max}$ (MK) & $\tau_{\rm CHeB}$ (yr) & $\tau_{\rm TPAGB}$ (yr) & $M_{\rm lost} ^{\rm final}$ ($\Msun$) & $y_{\rm 26Al}$ ($\Msun$) & \iso{26}Al/\iso{27}Al \\ 
        \hline
        $2\Msun$ & 0.441 & 0.503 & 0.531 & 2.8 & $1.21 \times 10^8$ & $2.61 \times 10^6$ & 0.730 & $3.4 \times 10^{-7}$ &  0.004 \\
        $3\Msun$ & 0.421 & 0.576 & 0.598 & 6.3 & $1.09 \times 10^8$ & $1.71 \times 10^6$ & 0.731 & $4.8 \times 10^{-7}$ & 0.004 \\
        $5\Msun$ & 0.779 & 1.006 & 0.863 & 75.4 & $2.30 \times 10^7$ & $3.52 \times 10^5 $ & 0.369 & $2.6 \times 10^{-6}$ & 0.009 \\
        $7\Msun$ & 1.208 & 1.494 & 0.962 & 92.4 & $8.84 \times 10^6$ & $1.66 \times 10^5$ & 0.159 & $2.9 \times 10^{-5}$ & 0.073 \\
        $8\Msun$ & 1.460 & 1.760 & 1.052 & 100.0 & $6.45 \times 10^6$ & $8.60 \times 10^4$ & 0.089 & $3.9 \times 10^{-5}$ & 0.089 \\
        \hline
        \multicolumn{10}{c}{Binary-star Monash models} \\
        Model & $M_{\rm c, RGBtip}$ ($\Msun$) & $M_{\rm c, PostCHeB}$ ($\Msun$) & $M_{\rm c,1TP}$ ($\Msun$) & $T_{\rm bce}^{\rm max}$ (MK) & $\tau_{\rm CHeB}$ (yr) & $\tau_{\rm TPAGB}$ (yr) & $M_{\rm lost} ^{\rm final}$ ($\Msun$) & $y_{\rm 26Al}$ ($\Msun$) & \iso{26}Al/\iso{27}Al \\ 
        \hline
        $5+2\Msun$ & 0.790 & 1.498 & 0.960 & 92.0 & $1.64 \times 10^7$ & $1.31 \times 10^5$ & 0.176 & - & - \\
        $5+3\Msun$ & - & - & 0.873 & 86.2 & - & $8.47 \times 10^5$ & 0.003 & $9.3\times 10^{-5}$ & 0.142 \\
		\hline
	\end{tabular}
\end{table*}

\subsubsection{Envelope mass increased during the HG}

\begin{figure} 
        \centering 
        \includegraphics[width=\columnwidth]{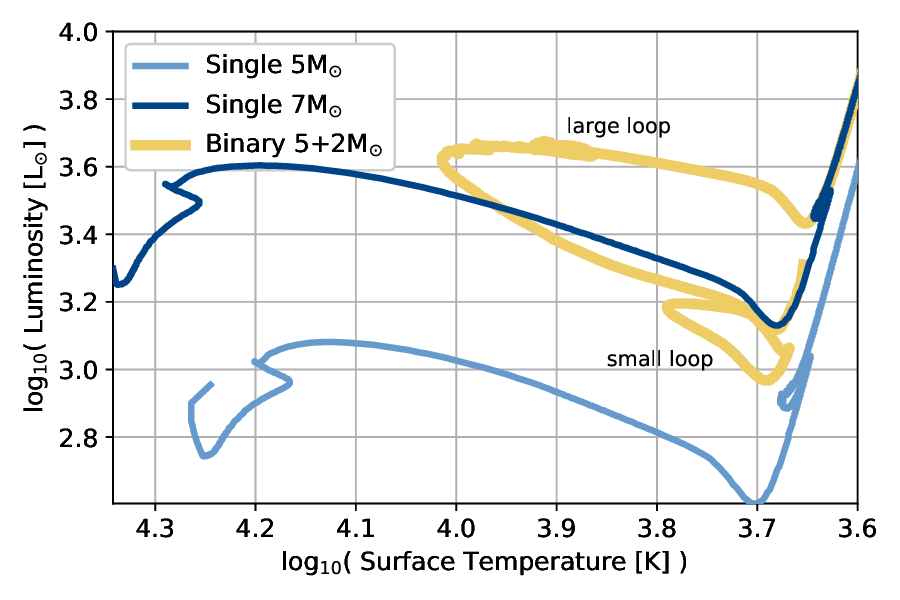}
        \caption[]{Hertzsprung-Russell diagram showing the evolutionary pathway of a single $5\Msun$ (light-blue) and single $7\Msun$ (dark-blue) stars from \emph{Monash16} in comparison to our $5+2\Msun$ star (yellow) after increasing the envelope mass post-MS. The "small loop" is from the relaxation of our stellar-model. Our $5+2\Msun$ stars ignites core helium with a $M_{\rm c}/M_{\rm tot} = 0.11$ and upon core He depletion $M_{\rm c}/M_{\rm tot} = 0.21$. $M_{\rm c}/M_{\rm tot}$ after core He depletion is identical to that of the single $7\Msun$ star after core He depletion. Our $5+2\Msun$ star experiences a CHeB phase with comparatively high surface temperatures compared to our $5\Msun$ and $7\Msun$ stars. Our $5+2\Msun$ star has a peak surface temperature of $10^{4.01} \, \mathrm{K}$ during CHeB causing the prominent blue loop (marked "large loop") observed in the HR diagram.}%
    \label{fig:HRDiagram}
\end{figure}

Fig. \ref{fig:HRDiagram} shows our $5+2\Msun$ (described in Section \ref{sec:MonashModels}) evolutionary sequences on the HR Diagram, compared to single-stars of $5\Msun$ and $7\Msun$. Our goal is to test if our $5+2\Msun$ star enters the TP-AGB with a core mass $\approx 0.86\Msun$, similarly to an identical star in \textsc{binary\_c}. Fig. \ref{fig:HRDiagram} shows that as a consequence of the increased envelope mass, our $5+2\Msun$ star experiences increased surface temperatures during CHeB with a peak surface temperature $T_{\rm eff,peak}$ of $10,200 \, \mathrm{K}$ which is over double compared to the $5\Msun$ ($T_{\rm eff,peak} = 4,700 \, \mathrm{K}$) and $7\Msun$ single-stars ($T_{\rm eff,peak} = 4,500 \, \mathrm{K}$. We refer to this behaviour as enhanced CHeB. Table \ref{tab:MonashModels} shows our $5+2\Msun$ star spends $1.68 \times 10^{7} \, \mathrm{yr}$ on the CHeB phase which is 1.7 times longer than our single $7\Msun$ star, but only 0.64 times as long as our single $5\Msun$ star. Our $5+2\Msun$ star enters the TP-AGB with a core mass of $0.96\Msun$ which is similar to that of a $7\Msun$ single-star from \emph{Monash16}. This star does not show similar behaviour to that of our overproducing stars in Fig. \ref{fig:OverAchieve}. Conditions for \iso{26}Al overproduction are not replicated in our $5+2\Msun$ star as the enhanced CHeB phase increases the core mass to that of a single $7\Msun$ star. As indicated in \ref{tab:MonashModels}, the single $5\Msun$ star produces $2.6\times 10^{-6} \Msun$ of \iso{26}Al and the single $7\Msun$ star produces over an order of magnitude more \iso{26}Al at $2.9\times10^{-5}\Msun$. Therefore, despite our $5+2\Msun$ star failing to enter the TP-AGB with a small core for its total mass, it would likely produce significantly more \iso{26}Al than the combined production of the single $5\Msun$ and $2\Msun$ stars.

\subsubsection{Envelope mass increased during the EAGB}

\begin{figure} 
        \centering 
        \includegraphics[width=\columnwidth]{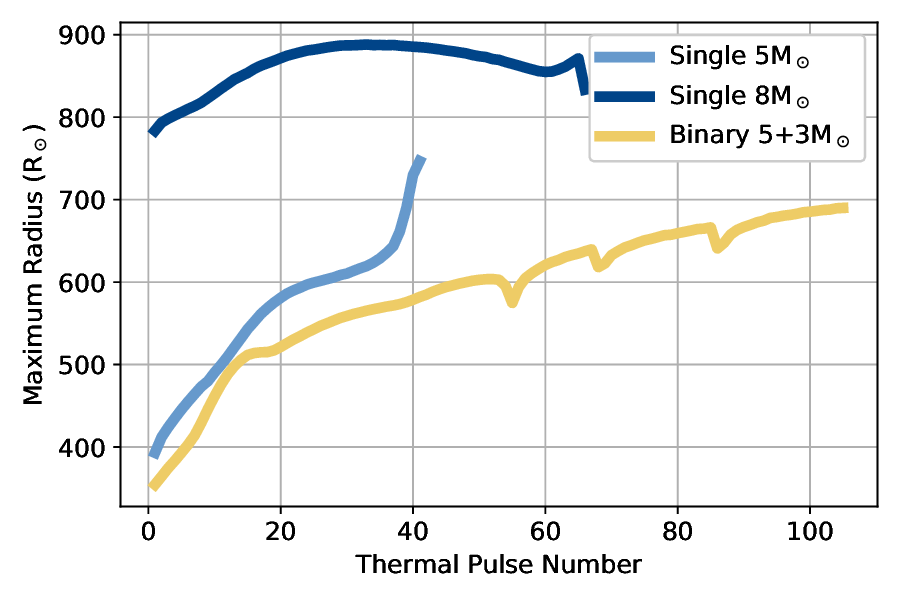}
        \caption[]{Maximum radius vs. thermal pulse count calculated by the Monash Models. Plot shows evolution of a single $5\Msun$ star (light blue), a single $8\Msun$ star (dark blue), and binary $5\Msun$ star which gains an additional $3\Msun$ during the EAGB (yellow). The increased envelope mass of our $5+3\Msun$ forces the convective envelope to shrink below that of the single $5\Msun$ star to sustain hydrostatic and thermal equilibrium.}%
    \label{fig:MM_Radius}
\end{figure}

\begin{figure} 
        \centering 
        \includegraphics[width=\columnwidth]{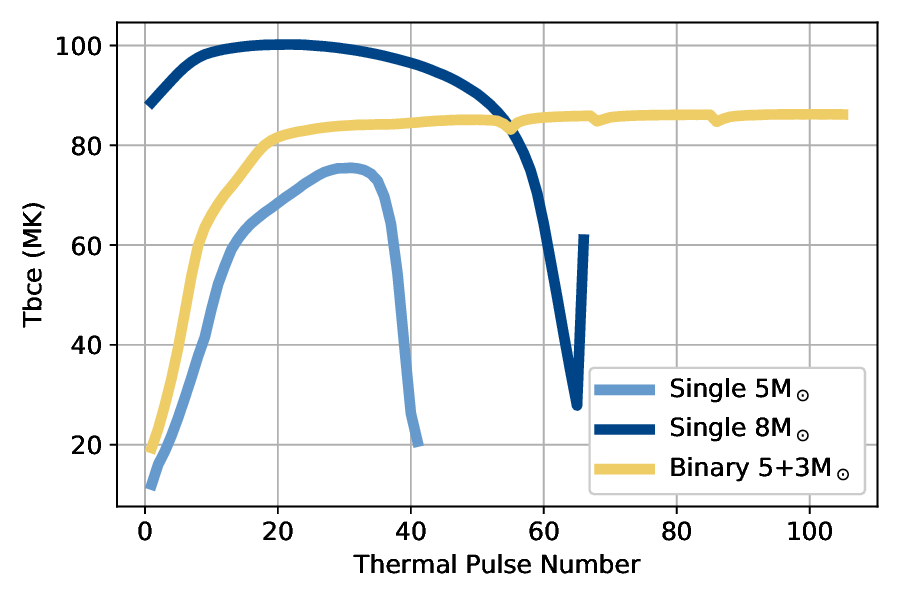}
        \caption[]{As Fig. \ref{fig:MM_Radius} but showing the temperature at the base of the convective envelope, $T_{\rm{bce}}$ vs. thermal pulse count as calculated by the Monash Models. Our $5+2\Msun$ star experiences hotter hot-bottom burning temperatures compared to the single $5\Msun$ star due to the additional $3\Msun$ of material in the envelope. Our $5+2\Msun$ star has a $M_{\rm c,1TP} = 0.87\Msun$ compared to the $M_{\rm c,1TP} = 1.05\Msun$ of the single $8\Msun$ star leading to lower HBB temperatures in our $5+3\Msun$ star compared to the single-star $8\Msun$ star. However our $5+3\Msun$ star experiences more thermal pulses and spends more time HBB compared to both the single $5\Msun$ and $8\Msun$ stars.}%
    \label{fig:MM_Tbce}
\end{figure}

\begin{figure} 
        \centering 
        \includegraphics[width=\columnwidth]{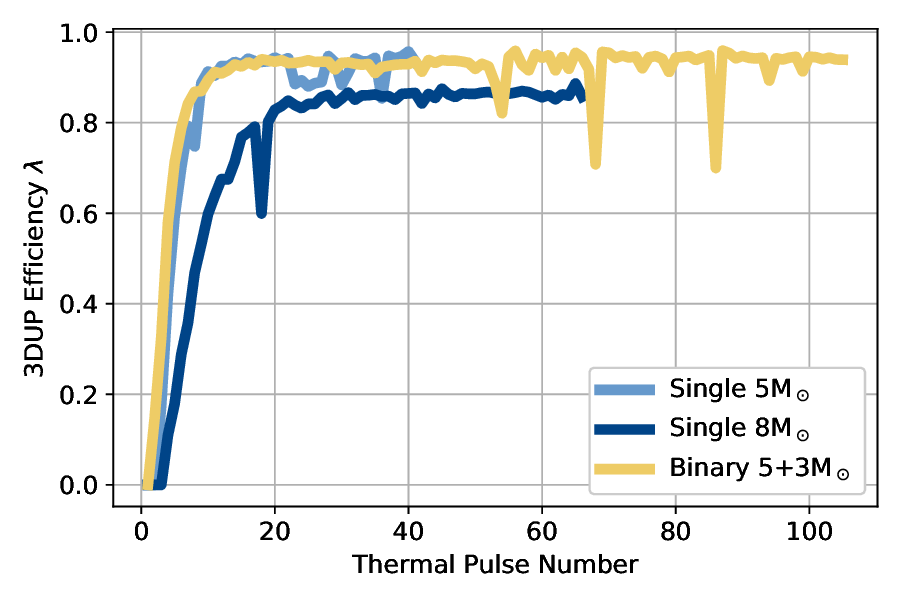}
        \caption[]{As Fig. \ref{fig:MM_Radius} but showing the efficiency of the third dredge up $\lambda$, vs. thermal pulse count as calculated by the Monash Models. We find our $5+3\Msun$ star has a similar $\lambda$ to the single $5\Msun$ star making the third dredge up in our $5+2\Msun$ than in the single $8\Msun$.}%
    \label{fig:MM_Lambda}
\end{figure}

\begin{figure} 
        \centering 
        \includegraphics[width=\columnwidth]{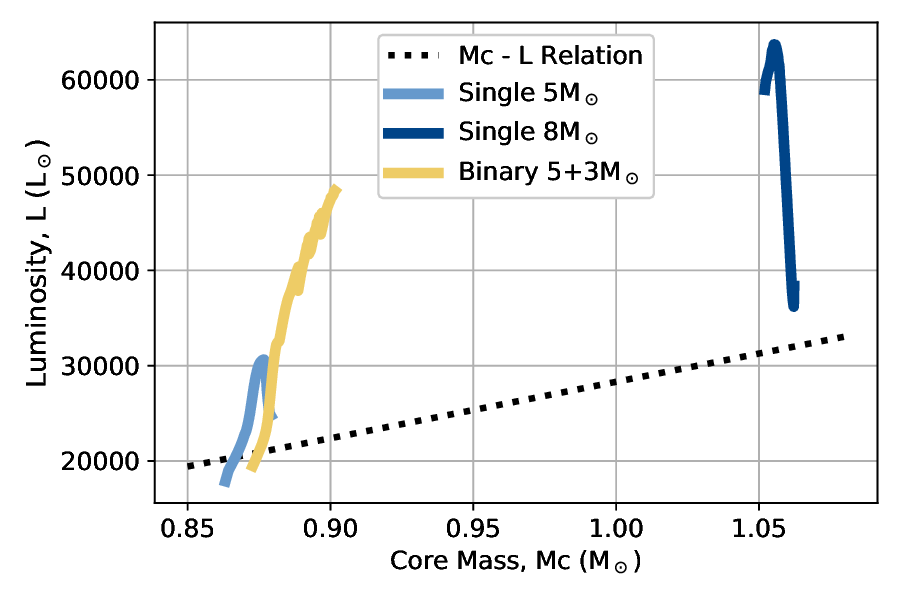}
        \caption[]{Core mass vs. Luminosity during the TP-AGB calculated by the Monash Models. The $M_{\rm c}-L$ relation is from \citet{Paczyncki1970}. Plot shows evolution of a single $5\Msun$ star (light blue), a single $8\Msun$ star (dark blue), and binary $5\Msun$ star which gains an additional $3\Msun$ during the EAGB (yellow). We find the increased envelope mass of our $5+3\Msun$ star compared to the single $5\Msun$ star leads to increased HBB temperatures (see Fig. \ref{fig:hbbtmax}) and increased luminosity. However our $5+3\Msun$ star does not becomes as luminous as the single $8\Msun$ star.}%
    \label{fig:CoreMassL}
\end{figure}

\begin{figure} 
        \centering 
        \includegraphics[width=\columnwidth]{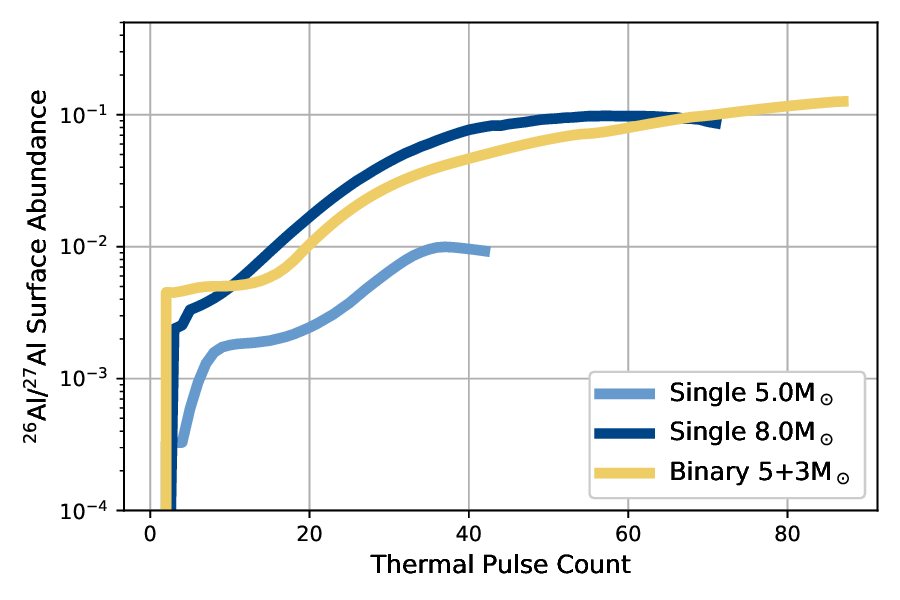}
        \caption[]{Surface \iso{26}Al/\iso{27}Al against thermal pulse number for a single $5\Msun$ star (light blue), a single $8\Msun$ star (dark blue), and our binary $5\Msun$ star which gains an additional $3\Msun$ during the EAGB (yellow). The \iso{26}Al/\iso{27}Al surface abundance ration of our $5+3\Msun$ star surpasses the peak \iso{26}Al/\iso{27}Al $= 9.82 \times 10^{-2}$ achieved by the $8\Msun$ star after 69 thermal pulses. Our $5+3\Msun$ star achieves a surface \iso{26}Al/\iso{27}Al = 0.13 at thermal pulse 86.}%
    \label{fig:SurfaceAl26K}
\end{figure}

Our $5+3\Msun$ Monash stellar-model as described in Section \ref{sec:MonashModels} has $3\Msun$ of additional mass dumped onto the envelope after the completion of core He-burning. See Table \ref{tab:MonashModels} for a summary of our results. Our $5+3\Msun$ star enters the TP-AGB with a total mass of $8.0\Msun$ and a core mass of $0.87\Msun$ which is almost identical to the core mass from the single $5\Msun$ solar metallicity star presented in \citet{Karakas2014_2}. We evolve our $5+3\Msun$ star on the TP-AGB phase through $8.47 \times 10^5$ yr and 105 thermal pulses. Our $5+3\Msun$ model sequence has not yet experienced the onset of the superwind when it ends but we estimate, based on a linear fit to the radial pulsation period between thermal pulse numbers $20-100$, this will occur at about thermal pulse number 139. At pulse 105 our $5+3\Msun$ star has spent 2.4 times longer on the TP-AGB than the single $5\Msun$ star.

Fig. \ref{fig:MM_Radius} shows the maximum stellar radius during each thermal pulse of our $5+3\Msun$ star compared to the single $5\Msun$ and $8\Msun$ stars modelled using the Monash code. Fig. \ref{fig:MM_Radius} shows that our $5+3\Msun$ star is smaller than even the single $5\Msun$ star. We speculate that the low radius is the cause of the increased gravitational pressure induced by the additional envelope mass forcing the star to shrink to achieve hydrostatic and thermal equilibrium. Convective envelopes are known to shrink with increased mass because $R \sim M^{-1/3}$ in fully convective polytropic stars \citep{Chandrasekhar1939}. The smaller radius of our $5+3\Msun$ star also results in a reduced mass loss rate as the Mira radial pulsation period depends on radius \citep{Vassiliadis1993}. Since our $5+3\Msun$ star sequence fails to evolve to the end of the TP-AGB, it is expected that the radius will continue to increase and probably surpass the maximum radius achieved by the single $5\Msun$ star.

Fig. \ref{fig:MM_Tbce} shows the temperature at the bottom of the convective envelope, $T_{\rm {bce}}$, vs. thermal pulse number in our $5+3\Msun$ star and the single $5\Msun$ and $8\Msun$ stars modelled using the Monash code. $T_{\rm{bce}}$ in our $5+3\Msun$ star is both increased compared to the single $5\Msun$ star and, unlike both the single $5\Msun$ and $8\Msun$ stars, is almost constant ($T_{\rm bce} \approx 86$ million K) through many thermal pulses. Fig. \ref{fig:MM_Tbce} also shows that our $5+3\Msun$ star does not surpass the peak $T_{\rm {bce}}$ of the $8\Msun$ star.

Fig. \ref{fig:MM_Lambda} shows the efficiency of third dredge up, $\lambda$, against thermal pulse number in our $5+3\Msun$ star and the single $5\Msun$ and $8\Msun$ stars modelled using the Monash code. We find our $5+3\Msun$ star has a similar $\lambda$ ($\lambda_{\rm max} = 0.96$) to the single $5\Msun$ ($\lambda_{\rm max} = 0.95$) rather than the single $8\Msun$ ($\lambda_{\rm max} = 0.89$). The increased $\lambda$ of our $5+3\Msun$ star aids in the production of \iso{26}Al by allowing the transport of more \iso{26}Al, that is not destroyed by neutron capture, to the surface via the third dredge up, albeit a small amount compared to the \iso{26}Al synthesized via HBB. We speculate that the increased efficiency of the third dredge up in our $5+3\Msun$ star, compared to our $8\Msun$ star, could potentially have implications for isotopes which rely on the third dredge up to be transported to the stellar surface and ejected, such as s-process elements.

Fig. \ref{fig:CoreMassL} presents the core mass vs. luminosity in our $5+3\Msun$ star, single $5\Msun$ and single $8\Msun$ stars modelled using the Monash code alongside the $M_{\rm c}-L$ relation from \citet{Paczyncki1970}. Fig. \ref{fig:CoreMassL} shows that our $5+3\Msun$ star has a similar core mass ($M_{\rm c,1TP} = 0.87\Msun$) to the $5\Msun$ star ($M_{\rm c,1TP} = 0.86\Msun$) but its luminosity exceeds the maximum luminosity of the single $5\Msun$ star and the $M_{\rm c}-L$ relation because of the hotter base of the convective envelope (Fig. \ref{fig:MM_Tbce}) driven by the increased compression of the relatively massive envelope. The peak luminosity of our $5+3\Msun$ star ($L_{\rm peak} = 4.83 \times 10^4 \Lsun$) does not surpass the peak of the $8\Msun$ star ($L_{\rm peak} = 6.37 \times 10^4 \Lsun$). 

Fig. \ref{fig:SurfaceAl26K} shows the \iso{26}Al/\iso{27}Al surface abundance ratio of the $5\Msun$, $8\Msun$, and our $5+3\Msun$ stars modelled using the Monash code against the thermal pulse count. Despite not achieving as high $T_{\rm {bce}}$ as that of the $8\Msun$ star, as shown in Fig. \ref{fig:MM_Tbce}, the sustained $T_{\rm {bce}}$ in our $5+3\Msun$ star causes it to surpass the $8\Msun$ surface \iso{26}Al/\iso{27}Al $= 9.82 \times 10^{-2}$ abundance ratio after 69 thermal pulses. We calculate the surface \iso{26}Al/\iso{27}Al ratio through 86 thermal pulses (when \iso{26}Al/\iso{27}Al $= 0.13$). We expect the surface \iso{26}Al/\iso{27}Al to continue climbing until the onset of the superwind near thermal pulse number 140. Taking the \iso{26}Al surface abundance at thermal pulse number 85 and the mass of the remaining envelope, we estimate that the \iso{26}Al stellar yield will exceed $9.3\times 10^{-5} \Msun$, at least 2.4 times higher than the single $8.0\Msun$ star. 
 
\section{Discussion}
\label{sec:Discussion}

This section explores the significance and limitations of our results and potential implications.

\subsection{Detailed vs synthetic models}
Our goal for the Monash models was to determine if we can replicate the small core and massive envelope conditions shown in Fig. \ref{fig:1TPMc} that allowed for \iso{26}Al overproduction in synthetic stellar-models, and to test if the Monash Models also experience prolonged HBB and \iso{26}Al overproduction compared to single-star models.

\subsubsection{Our $5+2\Msun$ star and potential implications}
As shown in Fig. \ref{fig:HRDiagram} our $5+2\Msun$ star modelled using the Monash code experienced enhanced CHeB with increased surface temperatures ($T_{\rm eff,peak} = 10,200 \, \mathrm{K}$), over double compared to our single $5\Msun$ and $7\Msun$ stars. Our $5+2\Msun$ star increased its core mass from $M_{\rm c,RGBtip} = 0.79\Msun$ (a similar $M_{\rm c,RGBtip}$ to our single $5\Msun$ star, see Table \ref{tab:MonashModels}) to $M_{\rm c,PostCHeB} = 1.49\Msun$ (similar $M_{\rm c,PostCHeB}$ to a single $7\Msun$ star). This behaviour is not reproduced in \textsc{binary\_c} and this is attributed to \textsc{binary\_c} relying on the stellar mass as the star begins to cross the HG ($M_{\rm PostMS}$) to calculate how the stars evolve post-MS (for example see Section \ref{sub:CoreMassEAGB}). This works for single-stars but does not take subsequent mass accretion, mass loss, or some mergers (such as the GB + MS star in Fig. \ref{fig:SurfaceAl26}) into account. If the duration of CHeB and core growth were modelled in \textsc{binary\_c} based on the total mass during CHeB, then 6295 out of 7297 (86.3\%) of our \iso{26}Al overproducing stars identified in Fig. \ref{fig:OverAchieve} might have more massive cores than indicated in Fig. \ref{fig:1TPMc}. This would mostly impact the "No Merger" cases (which make up 55\% of overproducing systems) identified in Fig. \ref{fig:OverAchieve} as these stars are HG, GB, or CHeB stars when they accrete. 

We do not explore the full implications of an enhanced CHeB phase on \iso{26}Al production. The enhanced CHeB experienced by our $5+2\Msun$ star may be a consequence of when we assumed the merger to happen. If we accrete the extra mass when the H-core was better established, or accrete mass more slowly, we may have found different behaviour. Our $5+2\Msun$ star enters the TP-AGB with similar core and envelope mass as our single $7\Msun$ star, and experiences similar HBB temperatures and duration. We estimate a our $5+2\Msun$ would have a similar yield as the single $7\Msun$ star of $2.9 \times 10^{-5}\Msun$ which is an order of magnitude higher than the combined yield of $2.9 \times 10^{-6} \Msun$ from our single $2\Msun$ and $5\Msun$ stars. A comprehensive study using detailed models is required to fully understand the enhanced CHeB and its consequences on stellar evolution and nucleosynthesis.

\textsc{binary\_c} uses stellar-timescales described in Section 5.3 of \citet{Hurley2000} to calculate models prior to the TP-AGB. Stellar parameters such as H-exhausted core mass, luminosity, and evolutionary phase are all dependent on these stellar-timescales. The stellar-timescales are fit to the \emph{Pols98} single-star models, however the enhanced CHeB phase is a consequence of binary evolution and is therefore outside the bounds of the fits. For \textsc{binary\_c} to adequately model enhanced CHeB, we would need to refit the stellar-timescales, luminosities, and radii etc. using binary detailed models. This result highlights the need to involve detailed binary models when constructing the tables and fitting formulae used by binary population synthesis \citep[e.g. see POSYDON,  ][]{Fragos2023}; single-star models are not sufficient. 




\subsubsection{Our $5+3\Msun$ star and mass accretion post-CHeB}
The $5+3\Msun$ star behaves like neither the $5\Msun$ nor $8\Msun$ stars. Fig. \ref{fig:CoreMassL} shows that the core mass of our $5+3\Msun$ star remains similar to that of our $5\Msun$ star but it is significantly more luminous due to a hotter HBB phase, similar to how \iso{26}Al overproducing stars evolve in \textsc{binary\_c}. We estimate that our $5+3\Msun$ star, which enters the TP-AGB with a total mass of $8.00\Msun$ and a H-exhausted core mass of $0.873\Msun$, has an \iso{26}Al yield of $> 9.3 \times 10^{-5} \Msun$ and our similar (if we ignore the possibility of an enhanced CHeB phase) \textsc{binary\_c} star shown in Fig. \ref{fig:SurfaceAl26}, which enters the TP-AGB with a total mass of $8.26\Msun$ and a H-exhausted core mass of $0.87\Msun$, has an \iso{26}Al yield of $2.0 \times 10^{-4}\Msun$. Additionally, overproducing stars modelled using \textsc{binary\_c} and Monash codes both have abnormally long TP-AGB phases (see Fig. \ref{fig:TpTime} and Table \ref{tab:MonashModels}). Our star modelled using \textsc{binary\_c} shown in Fig. \ref{fig:SurfaceAl26} spends $3.65 \times 10^{5} \, \mathrm{yr}$ on the TP-AGB, which is 2.3 times shorter than our $5+3\Msun$ star modelled using the Monash code, and 72 thermal pulses. We estimate our $5+3\Msun$ star will experience a total of 140 thermal pulses, almost double the thermal pulses from our star modelled using \textsc{binary\_c} in Fig. \ref{fig:SurfaceAl26}. Fig. \ref{fig:TpTime} show that a TP-AGB lifetime of $8.47\times10^5 \, \mathrm{yr}$ is longer than all of our \textsc{binary\_c} \iso{26}Al overproducers which enter the TP-AGB with $8\Msun$, and our $5+3\Msun$ star has not yet begun the superwind.

From the merger channels identified in Fig. \ref{fig:OverAchieve}, our $5+3\Msun$ most closely represents the "TP-AGB + HG" case, which results in an EAGB star post-merger. There are 6 \iso{26}Al overproducers from our $80 \times 80 \times 80$ grid evolving down this channel.

\subsection{Synthetic model uncertainty}
Detailed single-star models are subject to uncertainty regarding AGB evolution \citep{Busso1999, Herwig2005, Karakas2014} and mass loss \citep{Stancliffe2007, Hofner2018}. The uncertainty in the rate of \iso{26}Al destruction via the \iso{26}Al(p,$\gamma$) reaction has also been shown to alter \iso{26}Al yields in single-star models by a  factor of about 2 \citep{Izzard2007, Siess2008}. For this work we use the reaction rates from the Nuclear Astrophysics Compilation of Reaction Rates (NACRE) collaboration \cite{Angulo1999}. More recent reaction rates are available, e.g. see \citet{Zhang2023} for an update on the \iso{25}Mg(p,$\gamma$)\iso{26}Al reaction rate from the Jinping Underground Nuclear Astrophysics Experimental Facility (JUNA), or \citet{Straniero2013} from the Laboratory for Underground Nuclear Astrophysics (LUNA) collaboration. However, investigating their inpact is beyond the scope of the paper. Due to the approximate nature of synthetic models, these uncertainties are exacerbated. The inclusion of binary evolution introduces additional uncertainty with, chiefly the inclusion of mass transfer and common envelope evolution. A detailed study of the uncertainties involved in binary evolution is outside of the scope of this paper, so it is unclear at this time the significance of our results presented in Table \ref{tab:PopYieldFrac}.

\subsection{Eccentricity}

In this paper we explored how varying initial primary mass, secondary mass, and orbital period affect \iso{26}Al production in AGB stars. One key parameter which is not explored is the initial eccentricity. Table \ref{tab:Inputs} shows we set the initial eccentricity for all binary systems to zero. Systems with an initial orbital period of $\lesssim 10$ days are believed to circularize due to tidal interaction within the lifetime of the system \citep{Hurley2002}, however wider binaries with zero eccentricity have been rarely observed \citep{Duquennoy1991, Raghavan2010, Tokovinin2016}. An eccentric orbit in a wide binary would likely increase the possibility of a RLOF or CE event within the binary system and will need to be included for our results to better reflect observations. 

\subsection{Questionable results in \textsc{binary\_c}}

\subsubsection{Synthetic models with mass > $6.5\Msun$}
All synthetic stellar-models are calibrated to the \emph{Monash02} and \emph{Monash16} models for CO-core mass, maximum third dredge up parameter, TP-AGB luminosity, and temperature at the base of the convective envelope up to $8\Msun$ (see Section \ref{sec:PopSynth}). A notable update missing from our modified version of \textsc{binary\_c} is the HBB calibration table \citep[see Table A.3 in ][]{Izzard2006} which controls some nuclear burning properties during HBB such as how quickly temperatures at the bottom of the convective envelope reach $T_{\rm bce,max}$ (see Section \ref{sub:hbbtmax}) and the fraction of the convective envelope actively HBB. It was found in \citet{Izzard2004} that chemical abundances are extremely sensitive to small deviations in the HBB calibration table. Since ejected \iso{26}Al from AGB stars is primarily synthesized via HBB, an improvement to our method could be to extend the HBB calibration table in \textsc{binary\_c} up to $8\Msun$.

We calibrate our modified version of \textsc{binary\_c} up to $8\Msun$ based on available \emph{Monash16} models, however Fig. \ref{fig:Al26Single} shows that synthetic single-stars end their lives as supernovae after $8.30\Msun$. This means the single-stars of initial mass between $8\Msun$ and $8.30\Msun$ are uncalibrated for some stellar parameters such as radius and luminosity. \textsc{binary\_c} handles this region by holding stellar parameters constant after $8\Msun$. For example Eq. \ref{eq:LamMax} is dependent on stellar mass as it enters the HG but in our modified version of \textsc{binary\_c} it is capped at $8\Msun$ effectively holding $\lambda_{\rm {max}}$ constant at 0.89 for models of mass $>8\Msun$ \citep[based on results from ][ we do not expect $\lambda_{\rm max}$ to drastically change]{Doherty2015}. It is unclear if all the stellar parameters of our synthetic stars of masses $8.0-8.3\Msun$ are behaving similarly to stars calculated using detailed stellar-models.

\subsubsection{An interesting \iso{26}Al overproduction channel}
An interesting binary overproducer is the "TP-AGB + HG" channel. Fig. \ref{fig:1TPMc} shows that the TP-AGB + HG merger case results in a new star which may enter the TP-AGB with a total mass $>9\Msun$ and a core mass $<0.95\Msun$ at the first thermal pulse. Fig. \ref{fig:TpTime} indicates these stars spend more than 5 Myr on the TP-AGB. Two of these stars spend so long on the TP-AGB they burn all of the H in their envelopes via HBB, resulting in a naked-He star which later explodes as a stripped core collapse supernova. These stars are infrequent, with only four overproducers in our \textsc{binary\_c} $80 \times 80 \times 80$ population evolving into naked-He stars, but they gain an \iso{26}Al/\iso{27}Al surface abundance ratio of over 0.5. It would be interesting to look at how these systems behave with a detailed stellar evolution code, especially given that a total mass of $9\Msun$ lies outside \textsc{binary\_c} calibration for AGB evolution.

\subsection{\iso{26}Al contribution from supernovae}
Fig. \ref{fig:Evol7} shows that binary evolution can sometimes result in supernovae. We do not consider the yields originating from supernovae in this study. One major of source of \iso{26}Al underproduction shown in the top panel of Fig. \ref{fig:PopYields} from stars with an initial mass of at least $7.5\Msun$ is due to supernovae truncating stellar evolution. Models from \citet{Limongi2018} show that a core collapse supernova from a single $13\Msun$ star could produce an \iso{26}Al stellar yield on the order of $10^{-5} \Msun$, which is on the same order of magnitude as our synthetic single-stars of initial mass greater than about $7\Msun$ as presented in Fig. \ref{fig:Al26Single}. In our \textsc{binary\_c} grid of $80 \times 80 \times 80$, we have 65,876 electron-capture and core-collapse supernovae. The inclusion of supernovae is expected to increase the \iso{26}Al contribution from intermediate-mass stars and increase the overall influence from binary evolution.

\subsection{Comparison to observed \iso{26}Al abundances}
Presolar grains are grains of stardust which become encased within meteorites. This is separate from meteoric abundances which are thought to reflect the abundances upon which the solar system formed. Presolar grains form in the stellar winds of stars with Group II \citep[as defined in ][]{Nittler1997} oxide grains thought to originate from HBB AGB stars. Presolar grains are our best observational source of \iso{26}Al from AGB stars, as their abundances reflect the surface abundances of the parent stars at the time of ejection. Other observational sources of \iso{26}Al such as direct measurement from molecular clouds \citep{Kaminski2018} are limited, and gamma-ray observations lack the resolution to measure individual stellar sources \citep{Naya1998}.

The disadvantage of presolar grains is that their parent stars must be inferred from their observed abundances, and are hence uncertain. HBB is believed to prevent carbon-star formation \citep{Boothroyd1993}, therefore presolar grains from HBB-stars are believed to be oxygen-rich. Stellar models of HBB stars \citep{Lugaro2017} utilizing the reaction rates from the LUNA collaboration \citep{Bruno2016} have resulted in oxygen and \iso{26}Al/\iso{27}Al abundances similar to those found in Group II oxide presolar grains.

Group II oxide presolar grains reach an \iso{26}Al/\iso{27}Al ratio of $\sim0.1$ \citep{Nguyen2004}, which is a little lower than single highest \iso{26}Al/\iso{27}Al achieved by our synthetic models of 0.6. Note that presolar grains may be diluted by surrounding material hence lowering isotopic abundances \citep[e.g. see ][]{Nguyen2007}. To perform a comprehensive comparison to Group II presolar grains we will need data for oxygen isotopes, which we do not calculate in detail for our models. The oxygen isotopes will be studied in future work.

A common topic of discussion surrounding \iso{26}Al is the overabundance of \iso{26}Al detected within the early solar system as inferred through excess of its daughter nuclide, $^{26}$Mg, in meteorites \citep{Lee1977}. Multiple theories have been put forward including pollution from a nearby supernovae \citep{Cameron1977, Schiller2015} or stellar winds originating from a Wolf-Rayet \citep{Gaidos2009} or AGB star \citep{Wasserburg2006, Lugaro2012, Wasserburg2017, Parker2023}, and synthesis within the protostellar disk \citep{Gaches2020}. The solar system is estimated to have formed with an \iso{26}Al/\iso{27}Al abundance of about $5.3 \times 10^{-5}$ \citep{Jacobsen2008, Mishra2014, Liu2019}, but some meteoric abundances have also been estimated to be less than $2 \times 10^{-6}$ \citep{Makide2011}. The \iso{26}Al/\iso{27}Al abundances from our synthetic binary models manage to span over this range, but our single star models do not with a minimum non-zero \iso{26}Al/\iso{27}Al abundance ratio of $7.4 \times 10^{-6}$ from the stellar yields. This leads to some support for an AGB-binary origin, but material from the interstellar medium will not be pure AGB-star ejecta hence diluting AGB-star material and lowering the \iso{26}Al/\iso{27}Al ratio by some unknown factor.

The radioactive isotope \iso{60}Fe is commonly used to constrain the initial conditions of the solar system's progenitor star via the ratio \iso{26}Al/\iso{60}Fe. Our results do not currently include \iso{60}Fe or other commonly discussed radionuclides such as \iso{107}Pd and \iso{182}Hf \citep{Adams2010, Lawson2022, Trueman2022}, therefore we cannot comment on the prospect of binary influence contributing to the abundance of presolar \iso{26}Al. \iso{60}Fe and other radionuclides will be studied in future work.

\subsection{AGB binary evolution and globular cluster abundance anomalies}
Measured abundances of stars in globular clusters have been a source of both confusion and frustration for decades. Observed abundances show enhanced N, Na, and Al alongside depleted C, O and Mg \citep{Sneden1997, Lamb2015, Salgado2022} likely due to proton-capture nucleosynthesis or extra mixing. Massive binaries \citep{deMink2009}, supermassive stars \citep{Denissenkov2014}, deep mixing during on the GB \citep{Sweigart1979, Fujimoto1999}, birth compositions \citep{Cottrell1981, DAntona2007}, fast rotating massive stars \citep{Decressin2007}, and AGB stars \citep{Fenner2004, Karakas2006, DErcole2010, Ventura2011} have been investigated as a potential polluters for various abundance pattern anomalies. Could the extended HBB experienced by the \iso{26}Al overproducers be a source of pollution in Galactic globular clusters? Globular clusters today are observed to have relatively low binary fractions ($< 10\%$) \citep{Davis2008, Ji2015, Lucatello2015}, however their initial binary fractions are uncertain. N-body simulations are unhelpful, with some suggesting globular clusters may have been formed with binary fractions up to $100\%$ which dramatically reduces over time \citep{Ivanova2005}, while others find that they mostly retain their initial binary fractions \citep{Hurley2007}. The $5+3\Msun$ Monash model and \textsc{binary\_c} overproducers presented in this study show enhanced N and depleted O at solar metallicity, lending some credibility to the idea that prolonged HBB in intermediate-mass binary systems could potentially be a source of the anomalous abundances observed in globular clusters. This will be further explored in future work.

\section{Conclusions}
\label{sec:Conclusion}
In this study we investigate the influence of binary evolution on the \iso{26}Al yields of low to intermediate-mass stars, and on a stellar population. Using the binary population synthesis code \textsc{binary\_c}, we find that the inclusion of binary effects increases the total weighted \iso{26}Al yield by 18\% at a binary fraction of 0.5 compared to our population of only single-stars. A binary fraction of 0.75 increased the total weighted population yield by 25\%. Population synthesis results show that mergers with at least one evolved star and WRLOF onto a secondary HG, GB, or CHeB star are the main channels which lead to \iso{26}Al overproduction in low- and intermediate-mass stars. These channels lead to synthetic stars entering the TP-AGB phase with small cores for their total masses, allowing them to spend an extended time on the TP-AGB and thereby extending the period of HBB. An example is our binary system of primary-star mass $5.10\Msun$, secondary mass of $5.05\Msun$ and orbital period of $0.15 \mathrm{yr}$ which enters the TP-AGB with a total mass of $8.27\Msun$ and $M_{\rm c,1TP} = 0.87\Msun$ after a merger. This system has a peak \iso{26}Al/\iso{27}Al surface abundance ratio of 0.41, over 200 times higher than our single $5\Msun$ and over double that of even our single $8.27\Msun$ star. Conversely, we find that binary systems which experience RLOF or CE events may prevent stars from entering the TP-AGB phase or lead to insufficient mass for HBB, thereby leading to an underproduction of \iso{26}Al.

We calculate stellar-models using the Mt Stromlo/Monash Stellar Structure Program to attempt to replicate the conditions which allow stars modelled using \textsc{binary\_c} to overproduce \iso{26}Al by increasing the envelope mass of a single $5\Msun$ star during the HG and EAGB phases. Our $5+2\Msun$ star, which gains $2\Msun$ during while crossing the HG, experiences an enhanced core He-burning phase. This star enters the core He-burning phase with a core mass similar to a single $5\Msun$ star but has a similar core and envelope mass to a single $7\Msun$ star at the first thermal pulse. This behaviour is not replicated by our stars modelled using \textsc{binary\_c} as 86\% of \iso{26}Al overproducing stars undergo CHeB after a merger and enter the TP-AGB with small cores for their total masses. A more thorough investigation with detailed models is required to fully understand the implications of an enhanced CHeB phase on our results. However we find our $5+2$ star behaves similarly to a single $7\Msun$ star upon entering the TP-AGB and will likely produce a similar \iso{26}Al yield resulting in an overall overproduction compared to the combined single $2\Msun$ and $5\Msun$ cases. 

Our $5+3\Msun$ star gains $3\Msun$ onto its envelope during the EAGB but before the second dredge up. Our $5+3\Msun$ star does enter the TP-AGB with a small core for its total mass and experiences a prolonged TP-AGB phase $\sim 10$ times longer than the single $8.0\Msun$ and $\sim 4$ times longer than the single $5\Msun$ star modelled using the Monash code. This leads to an extension of the TP-AGB phase and HBB resulting in an overproduction of \iso{26}Al compared to a single $5\Msun$ or $8\Msun$ star.

Our results show that \iso{26}Al yields of individual systems can be extremely sensitive to binary evolution leading to an overall overproduction in a stellar population as compared to a population of only single-stars. Further investigation involving detailed models including core growth during core He-burning is required to verify our synthetic models. Our results introduce the possibility of binary evolution being responsible for the anomalous abundances observed in globular clusters or for the overabundance of \iso{26}Al in our solar system. Our results show considerable promise in understanding the contribution from binary evolution onto the stellar yields from a low- and intermediate-mass stellar population.

\section*{Acknowledgements}

ZO acknowledges this research was supported by an Australian Government Research Training Program (RTP) Scholarship. AK and ZO were supported by the Australian Research Council Centre of Excellence for All Sky Astrophysics in 3 Dimensions (ASTRO 3D), through project number CE170100013.

RGI thanks STFC for funding \textsc{binary\_c} through grants ST/L003910/1 and ST/R000603/1, and acknowledges the support of the BRIDGCE consortium.

\section*{Data Availability}

Data can be made available upon reasonable request to the corresponding authors.



\bibliographystyle{mnras}
\bibliography{References}




\appendix


\bsp	
\label{lastpage}
\end{document}